\documentclass[acmtog,nonacm]{acmart}

\usepackage{booktabs} 
\usepackage{caption}
\usepackage{subcaption}
\usepackage{multirow}
\usepackage{enumitem}
\usepackage{engord}
\usepackage{microtype}
\usepackage{amsmath}
\usepackage{xspace}

\usepackage[ruled]{algorithm2e} 

\SetAlFnt{\small}
\SetAlCapFnt{\small}
\SetAlCapNameFnt{\small}
\SetAlCapHSkip{0pt}

\usepackage{soul}

\setcopyright{rightsretained}
\acmJournal{TOG}
\acmYear{2023}
\acmVolume{1}
\acmNumber{1}
\acmArticle{1}
\acmMonth{1}
\acmPrice{}
\acmDOI{10.1145/3618342}

\usepackage{ifthen}
\usepackage{soul}

\newcommand{\ie}{\emph{i.e.}}

\newcommand{\loss}{\mathcal{L}}
\newcommand{\TIspace}{\mathcal{P}}
\newcommand{\STIspace}{\mathcal{P}^{*}}

\newcommand{\sysname}{\textit{ProSpect}\xspace}

\usepackage{ifthen}
\definecolor{WeimingColor}{rgb}{0,0,0.8} 
\definecolor{FanColor}{rgb}{0.8,0,0.8}
\definecolor{OliverColor}{rgb}{0.8,0,0.8}
\newcommand{\weiming}[1]{{\color{WeimingColor} [Weiming: #1]}}
\newcommand{\fan}[1]{{\color{FanColor}[Fan: #1]}}

\newcommand{\revision}[1]{{\color{orange} #1}}
\newcommand{\reviseagain}[1]{{\color{cyan} #1}}
\newcommand{\delete}[1]{{\color{orange} \st{#1}}}

\newcommand{\final}{1}

\ifthenelse{\equal{\final}{1}}
{
\renewcommand{\weiming}[1]{}
\renewcommand{\fan}[1]{}

\renewcommand{\revision}[1]{#1}
\renewcommand{\reviseagain}[1]{#1}
\renewcommand{\delete}[1]{}
}
{}

\begin{document}
\emergencystretch=0.3em

\title{ProSpect: Prompt Spectrum for Attribute-Aware Personalization of Diffusion Models}

\author{Yuxin Zhang}
\affiliation{
\institution{MAIS, Institute of Automation, CAS}
\country{China}
}
\affiliation{
\institution{School of Artificial Intelligence, UCAS}
\city{Beijing}
\country{China}
}
\author{Weiming Dong}
\affiliation{
\institution{MAIS, Institute of Automation, CAS}
\country{China}
}
\affiliation{
\institution{School of Artificial Intelligence, UCAS}
\city{Beijing}
\country{China}
}
\author{Fan Tang}
\affiliation{
\institution{Institute of Computing Technology, CAS}
\city{Beijing}
\country{China}
}
  
\author{Nisha Huang}
\affiliation{
\institution{School of Artificial Intelligence, UCAS}
\country{China}
}
\affiliation{
\institution{MAIS, Institute of Automation, CAS}
\country{China}
}
 
\author{Haibin Huang}
\author{Chongyang Ma}
\affiliation{
\institution{Kuaishou Technology}
\city{Beijing}
\country{China}
}

\author{Tong-Yee Lee}
\affiliation{
\institution{National Cheng-Kung University}
\city{Tainan}
\country{Taiwan}
}

\author{Oliver Deussen}
\affiliation{
\institution{University of Konstanz}
\city{Konstanz}
\country{Germany}
}
\author{Changsheng Xu}
\affiliation{
\institution{MAIS, Institute of Automation, CAS}
\country{China}
}
\affiliation{
\institution{School of Artificial Intelligence, UCAS}
\country{China}
}
\renewcommand\shortauthors{Zhang et al.}

%
%



\begin{abstract}
Personalizing generative models offers a way to guide image generation with user-provided references.
Current personalization methods can invert an object or concept into the textual conditioning space and compose new natural sentences for text-to-image diffusion models.
However, representing and editing specific visual attributes such as material, style, and layout remains a challenge, leading to a lack of disentanglement and editability.
To address this problem, we propose a novel approach that leverages the step-by-step generation process of diffusion models, which generate images from low to high frequency information, providing a new perspective on representing, generating, and editing images. 
We develop the Prompt Spectrum Space $\STIspace$, an expanded textual conditioning space, and a new image representation method called \sysname.
\sysname represents an image as a collection of inverted textual token embeddings encoded from per-stage prompts, where each prompt corresponds to a specific generation stage (i.e., a group of consecutive steps) of the diffusion model.
Experimental results demonstrate that $\STIspace$ and \sysname offer better disentanglement and controllability compared to existing methods.
We apply \sysname in various personalized attribute-aware image generation applications, such as image-guided or text-driven manipulations of materials, style, and layout, achieving previously unattainable results from a single image input without fine-tuning the diffusion models.
Our source code is available at \url{https://github.com/zyxElsa/ProSpect}.

\end{abstract}

\begin{teaserfigure}
    \centering
    \includegraphics[width=\linewidth]{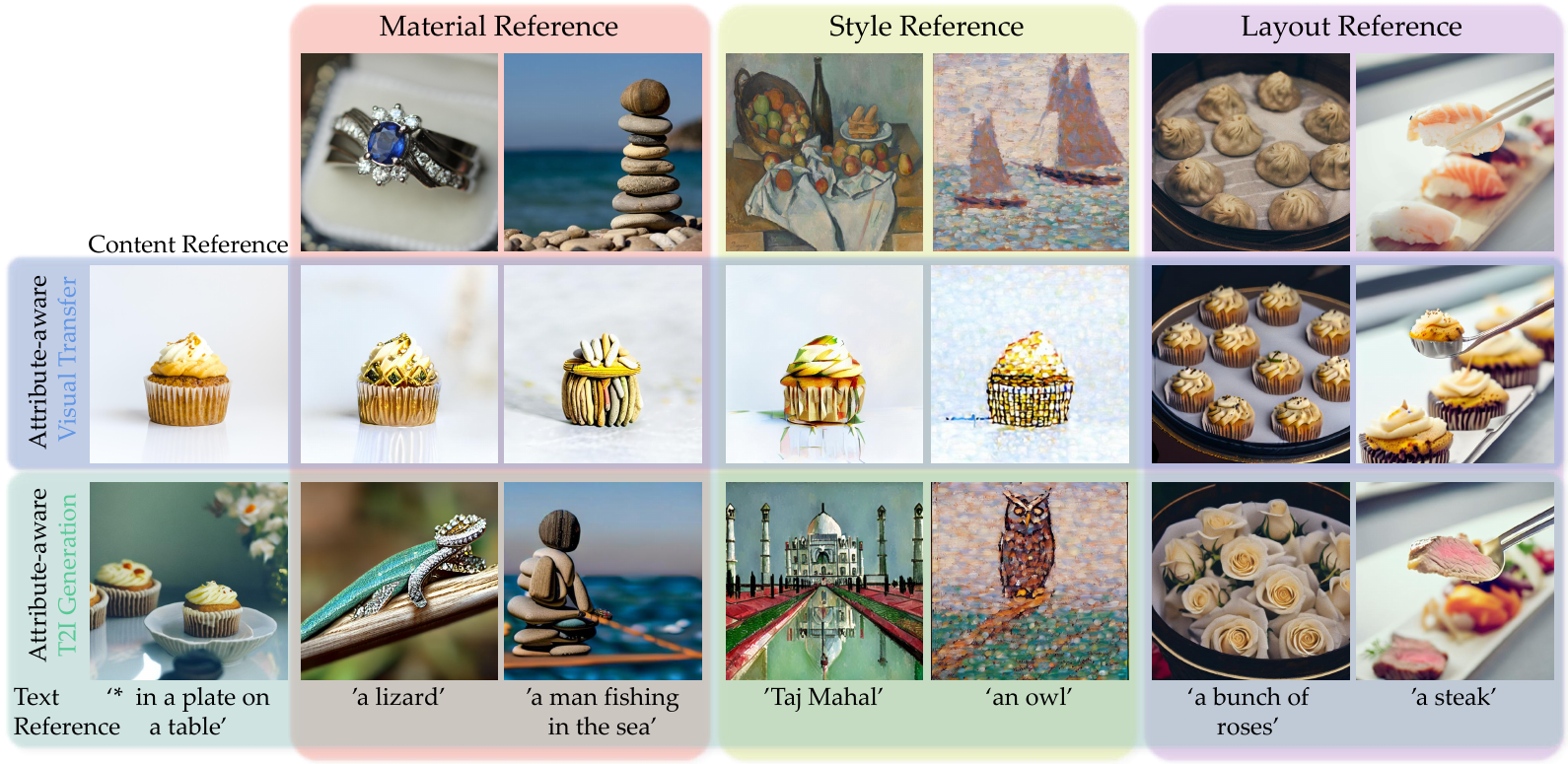}
    \caption{
    Attribute-aware image generation results using \sysname.
    Given a single input image or text prompts, our method can intuitively control visual attributes such as material, style, content, and layout to generate a new image with the learned textual conditionings.
    Real image credits (from left to right): \{Vojtech Okenka, Taisuke usui, Pixabay\}/Pexels (Free to use)~\cite{pexels}, Paul Cezanne/The Art Institute of Chicago (CC0)~\cite{AIC}, Georges Seurat/The Barnes Foundation (CC0)~\cite{barnes}, \{Rov Camato, Chevanon Photography\}/Pexels (Free to use)~\cite{pexels}.}
    \label{fig:teaser}
\end{teaserfigure}

\maketitle

\section{Introduction}
\label{sec:introduction}

If we consider photography and painting as visual languages, we can understand that each image encapsulates a unique perspective or way of seeing. By harnessing the power of pre-trained diffusion models designed for text-to-image generation, we obtain a versatile method for influencing the synthesis process using natural language commands. The utilization of these advanced generative models not only allows for the creation of realistic and diverse images but also enables users to personalize the output according to their visual preferences.
Recent personalization methods~\cite{gal2022TI,ruiz2022dreambooth,kumari2022customdiffusion,huang2023reversion} learn the textual conditioning of a common concept from a set of images and then use text prompts to create new scenarios that incorporate the concept.
However, representing specific visual attributes of a single image remains a challenging problem for these concept-level personalization methods.

\begin{figure}
\centering
\includegraphics[width=1\linewidth]{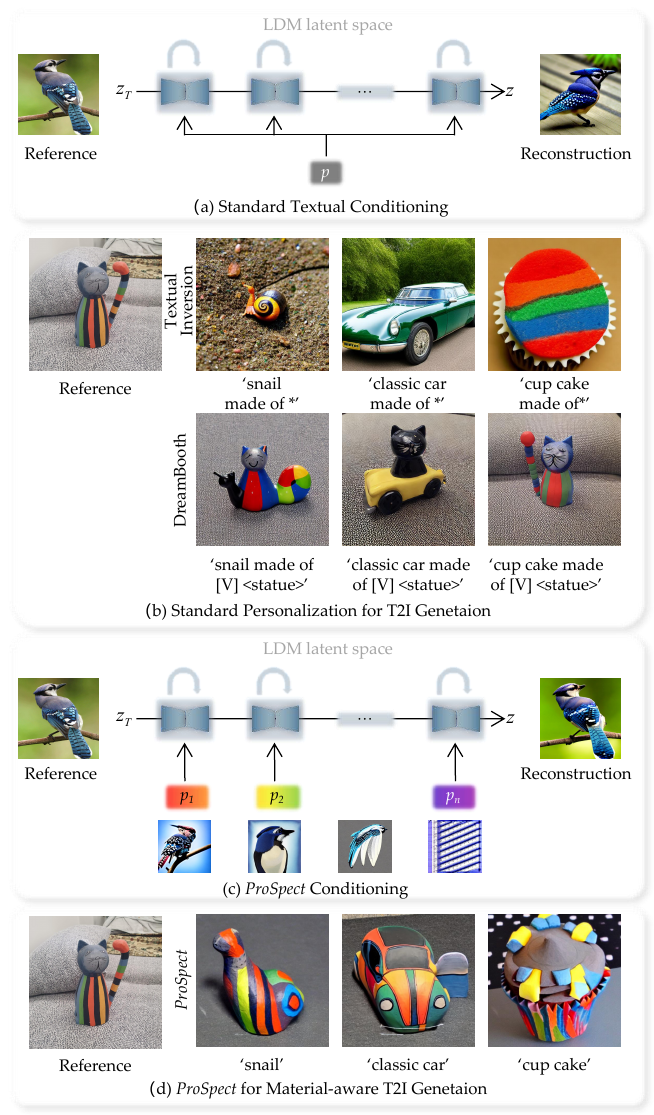}
\caption{Differences between (a) standard textual conditioning in $\TIspace$ and (c) prompt spectrum conditioning in $\STIspace$. Instead of learning global textual conditioning for the whole diffusion process, \sysname obtains a set of different token embeddings delivered from different denoising stages.
As shown in (b) standard personalization for T2I attribute-aware image generation, Textual Inversion~\cite{gal2022TI} loses some of the fidelity, and DreamBooth~\cite{ruiz2022dreambooth} generates cat-like objects in the images. (d) \sysname for attribute-aware generation shows that \sysname can separate content and material, and is more fit for attribute-aware T2I image generation.
Reference image credit: Pixabay/Pexels (Free to use)~\cite{pexels}.}
\label{fig:intro}
\end{figure}

We believe that each visual attribute (e.g., style, material, layout, etc.) within an image has its own unique features.
Attribute-aware image generation, therefore, involves the representation, disentanglement, and recombination of these visual attributes to guide image synthesis and editing.
The primary challenge lies in disentangling the specific attributes of a single image, as they often appear in combination.
Additionally, recombining the attributes without causing conflicts or distortions is difficult when performing image attribute transfer tasks.
By projecting image references into a conditioned textual space (defined as $\TIspace$ in Gal~\shortcite{gal2022TI}, see Fig.~\ref{fig:intro}(a)), text-to-image generation methods can conduct concept-level image editing.
However, generating single textual embedding across all diffusion steps and U-Net structures limits the ability for visual attribute disentanglement.   
In line with \citet{gal2022TI}, \citet{voynov2023pplus} observe that the shallow layers of the denoising U-Net structures within diffusion models tend to generate colors and materials, while the deep layers provide semantic guidance.
In this work, we conduct a detailed analysis of how textual conditioning influences the generation process of diffusion models.
We present various visualization results to demonstrate that diffusion models generate images in the order of \textit{layout $\rightarrow$ content $\rightarrow$material/style}.
Our further analysis reveals that the generation order in a diffusion model is correlated to the signal frequency of the corresponding attribute, which is progressed from low to high.
This insight paves the way for obtaining better disentanglement of visual attributes in diffusion models.

Inspired by this observation, we introduce Prompt Spectrum Space $\STIspace$ (see Fig.~\ref{fig:intro}(c)), an expanded conditioning space of $\TIspace$ that provides a new insight on the diffusion generation process from the perspective of \textit{steps}.
Instead of treating all diffusion steps as a whole, we consider several groups of consecutive steps as different generation \textit{stages}.
Each stage corresponds to a unique textual condition $p_i$.
We further propose a novel inversion and condition method \sysname, which learns token embeddings $P$ in $\STIspace$ from a single image.
Unlike previous methods that consider the concept or image as a whole, \sysname provides a new way to represent an image in the perspective of frequency, which improves flexibility and editability.
Various visual attributes can be separated from $P$, enabling attribute-aware generation.
Specifically, we group the textual token embeddings $p_i$ into three classes, \ie, material/style (high-frequency), content (medium-frequency), and layout (low-frequency).
By replacing them with embeddings of other images, we can achieve attribute transfer, as shown in the \engordnumber{2} row of Fig.~\ref{fig:teaser}.
Compared to previous personalization approaches, \sysname offers better transferability of diverse image visual attributes.
Notably, in the context of attribute-aware image-to-text generation tasks, \sysname demonstrates superior editability and fidelity, achieving results that were previously difficult to obtain, as shown in the \engordnumber{3} row of Fig.~\ref{fig:teaser}.
Figs.~\ref{fig:intro}(b) and \ref{fig:intro}(d) show the differences between different personalization methods applying to material controlling tasks, including Textual Inversion~\cite{gal2022TI}, DreamBooth~\cite{ruiz2022dreambooth}, and our \sysname.
Textual Inversion loses most of the fidelity.
Due to the lack of separation of content and material, DreamBooth tends to generate cat-like objects in each image.
\sysname separates content and material in the learning and conditioning process and can generate a new image that is only loosely related to the content of the reference image.
Extensive experiments and evaluations demonstrate the effectiveness of $\STIspace$ and \sysname.

To summarize, our contributions are:
\begin{itemize}[leftmargin=*]
    \item We introduce a novel Prompt Spectrum Space $\STIspace$ that enables the disentanglement of visual attributes from a single image.
     We also reveal that the generation process of diffusion models depends on the frequency of visual signals.
    \item We present Prompt Spectrum (\sysname), a novel image representation and manipulation method that offers better controllability and flexibility when processing visual attributes.
    \item Our experimental results demonstrate the effectiveness of $\STIspace$ and \sysname in various attribute-aware image generation tasks.
\end{itemize}

\section{Related Work}
\label{sec:relatedworks}

\paragraph{Text-to-image synthesis}
Generative Adversarial Network (GAN)-based architectures~\cite{Goodfellow:2014:GAN} are widely used in text-to-image models, which are trained on large sets of paired image-caption data~\cite{xu2018attngan,zhu2019dm,zhang2021cross,liao2022text,Tao:2022:dfgan}.
However, GANs have a tendency to suffer from mode collapse and their training at scale can be challenging~\cite{heusel2017gans,brock2019large}.
Auto-regressive models~\cite{gafni2022make,ramesh2021zero,yu2022scaling} are inspired by the success of language models and perform the task of image generation by treating images as word sequences in a discrete latent space~\cite{esser2021taming}.
This scheme allows for text guidance during generation through conditioning on text-prefix or using text-to-image similarity models~\cite{crowson2022vqgan,CLIPstyler,StyleGANNADA} at test-time optimization.
Recently, diffusion models~\cite{dhariwal2021diffusion,IDDPM} have emerged as the forefront of image generation. These models have led to significant advances in text-to-image synthesis, achieving more natural results with impressive diversity and fidelity~\cite{balaji2022ediffi,nichol2022glide,dalle2,latentdiffusion,saharia2022photorealistic,Huang2022MGAD,chang2023muse}.

\paragraph{Personalization of generative models}
The personalization of the text-to-image generation model is the task of generating personalized content based on the pre-trained model.
\citet{gal2022TI} present a textual inversion method to find a pseudo-word to describe the visual concept of a specific object.
\citet{gal2023encoder} further design a word-embedding encoder to predict a new pseudo-word that best describes the input concept.
\citet{li2023stylediffusion} invert the real image to the linear mapping network in cross-attention layers.
\citet{ruiz2022dreambooth} implant a subject into the output domain of a text-to-image diffusion model to synthesize it in novel views with a unique identifier.
\citet{zhang2023inst} propose an attention-based inversion style transfer method called InST.
\citet{kumari2022multi} propose Custom Diffusion, which optimizes a few parameters in the conditioning mechanism and can jointly train for multiple concepts or combine several fine-tuned models.
\citet{huang2023reversion} propose ReVersion for relation inversion, which aims to learn a specific relation from images.
\citet{wen2023hard} introduce the concept of hard prompts that use hand-crafted sequences of interpretable tokens to elicit model behaviors.
\citet{voynov2023pplus} present an extended textual conditioning space $\TIspace +$ that consists of multiple textual conditions, derived from per-layer prompts, each corresponding to a layer of the denoising U-Net of the diffusion model.
\citet{Tewel:2023:perfusion} introduce Perfusion, a mechanism that locks cross-attention keys of new concepts to their superordinate category, and a gated rank-1 approach to control the influence of a learned concept.

Most of the aforementioned methods necessitate an image set (three to five) as input or require model fine-tuning, and they aim to learn a single concept in the image or represent the overall appearance of the image.
In contrast, our approach addresses the challenges of obtaining multiple visual attributes from a single image, involving the representation, disentanglement, and recombination of visual attributes.

\begin{figure*}
\centering
\includegraphics[width=1\linewidth]{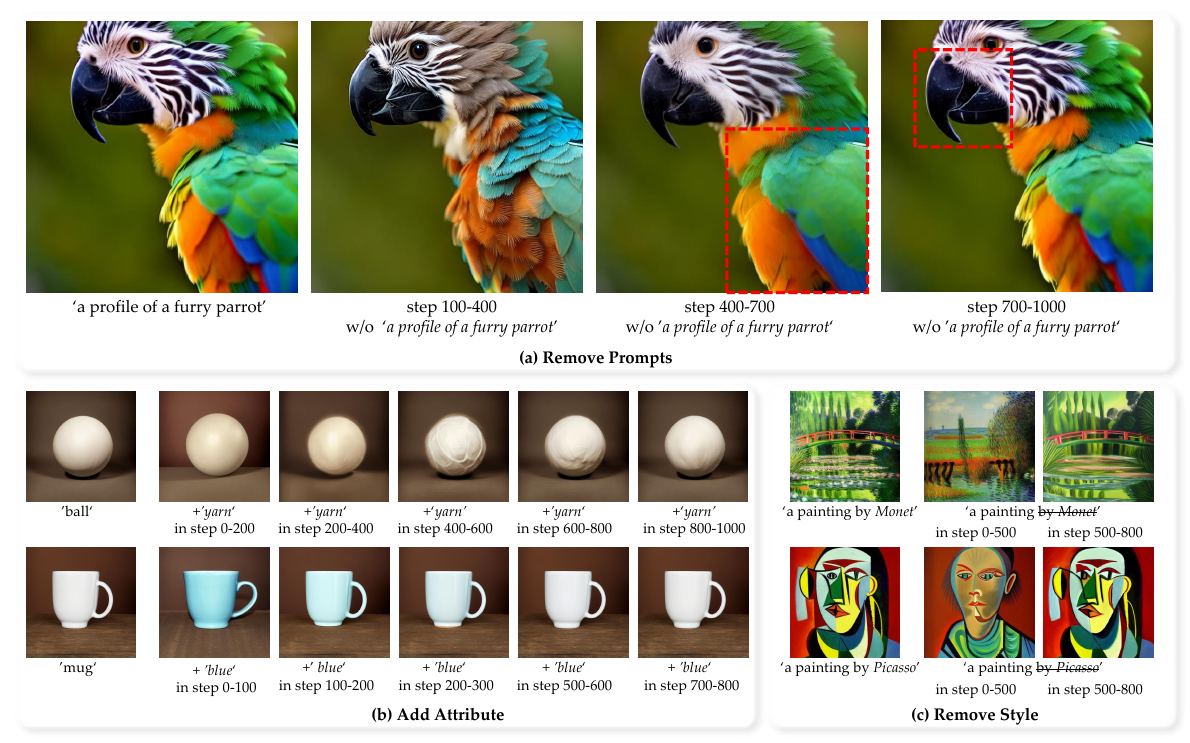}
\caption{Experimental results showing that different image attributes correspond to different generation steps.
(a) Results of removing prompts ``\textit{a profile of a furry parrot}'' of different steps.
(b) Results of adding material attribute ``\textit{yarn}'' and color attribute ``\textit{blue}''.
(c) Results of removing style attributes ``\textit{Monet}'' and ``\textit{Picasso}''.}
\label{fig:motivation}
\end{figure*}

\begin{figure}
\centering
\includegraphics[width=1\linewidth]{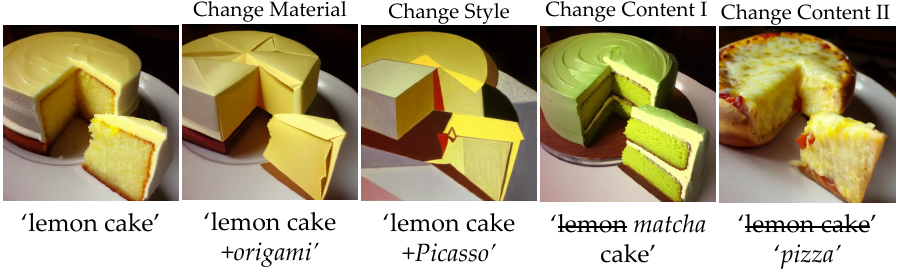}
\caption{Prompt-based editing results.
By changing the prompts conditioning on different diffusion stages and keeping the layout-related prompts unchanged, we can achieve the effect of prompt-to-prompt editing.
}
\label{fig:prompt_edit}
\end{figure}

\begin{figure*}
\centering
\includegraphics[width=1\linewidth]{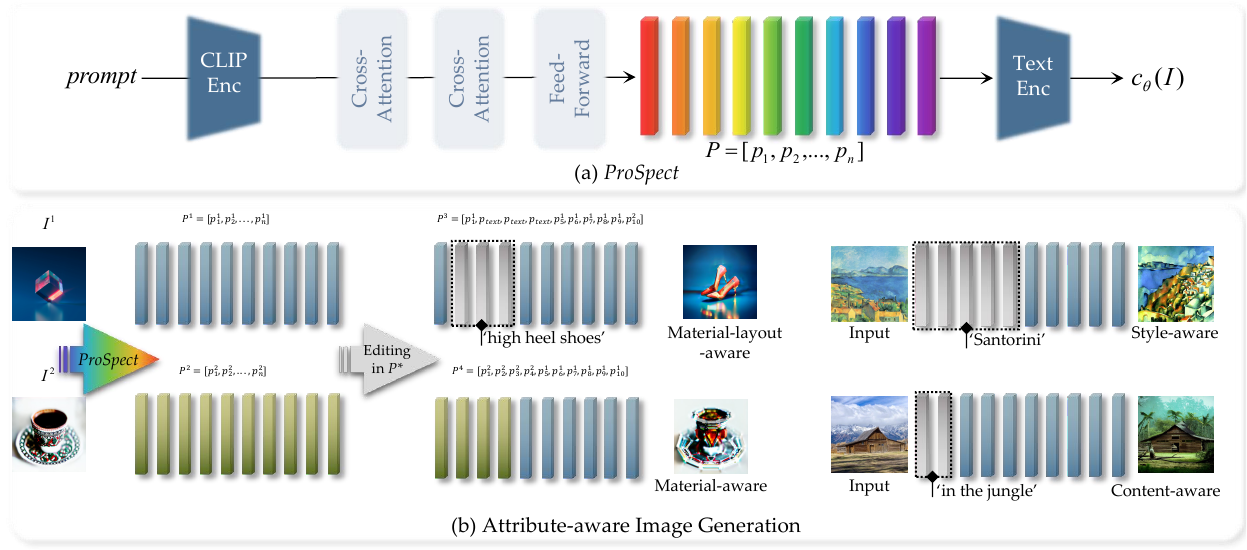}
\caption{
(a) The pipeline of \sysname, which learns a set of token embeddings $P=\left[p_1,p_2,...,p_n\right]$. 
(b) Illustrations of various attribute-aware image generation tasks.
Reference image credits: \{Rostislav Uzunov, Lisa Fotios\}
/Pexels (Free to use)~\cite{pexels}.
Style image credit (the \engordnumber{1} row): Paul Cezanne/The Art Institute of Chicago (CC0)~\cite{AIC}. }
\label{fig:pipeline}
\end{figure*}

\paragraph{Image editing}
A variety of text-based image editing methods~\cite{bau2021paint,StyleCLIP,StyleCLIPDraw} have emerged with the development of powerful multi-modal models.
Enabled by diffusion models, approaches of different applications are developed, such as single-image editing~\cite{brooks2022instructpix2pix,imagic,meng2022locating,meng2021sdedit,mokady2023null,zhang2023sine,valevski2022unitune,wu2022uncovering,huang2023region}, style transfer~\cite{jeong2023training,yang2023zero,huang2022diffstyler,huang2023style} and inpainting~\cite{avrahami2022blended,yang2022paint,lugmayr2022repaint}.
The Composer approach \cite{huang2023composer} is most relevant to our work.
This approach introduces a generation paradigm that enables control over the output features, while preserving synthesis quality and model creativity through decomposing images into representative factors (e.g., spatial layout and color palette) and training a diffusion model using these factors as conditions for recomposition.
However, they rely on additional task-specific models to obtain image attributes, such as an edge detection model for contour extraction, a pre-trained segmentation model for extraction of instances and the corresponding masks, etc.
In contrast, we exclusively use a pre-trained diffusion model to obtain the representation of corresponding attributes from the input image, which provides a neat way to disentangle and control visual attributes.

\revision{
Many non-diffusion image editing methods encode images into a latent space~\cite{Wang:2023:THR,Wang:2023:DFE,lee2020drit, Zhang:2023:UCAST}.
StyleGAN~\cite{stylegan} consists of a mapping network, which maps latent codes to the latent space $\mathcal{W}$, and a synthesis network, which controls the feature statistics between different network layers. Fine-grained control over semantic attributes in generated images is achieved by manipulating different dimensions of the latent vectors.
With the ability of generating high resolution images of high quality, StyleGAN and its followups ~\cite{karras2020training,StyleGANNADA} have become the advanced unconditional image generators.
FineGAN~\cite{singh2019finegan} disentangles the background, object shape,and object appearance to hierarchically generate images of fine-grained object categories. 
MUNIT~\cite{munit} decomposes the image into a domain-invariant content code and a style code that captures domain-specific properties, and achieves editing by recombining the codes.
SwappingAutoencoder~\cite{park2020swapping} encodes an image into two independent components and enforce that any swapped combination maps to a realistic image.
Differently, our approach encodes image attributes into the target text space and represents attributes separately using different embeddings.
Besides, the above latent space traversal is usually limited to editing within domains, in contrast, our method enables cross-domain editing.
}

\section{Method}
\label{sec:method}

To illustrate our motivation, we start by analyzing the attribute distribution of diffusion models using text-guided image generation results.
We aim to obtain multiple visual attributes from a single image, thus we need to learn the range of the steps in which different attributes are generated by the model.

Fig.~\ref{fig:motivation} shows the results of removing or adding attributes at different diffusion stages.
In Fig.~\ref{fig:motivation}(a), removing a certain phase ``a profile of a furry parrot'' in some steps will cause certain changes to the generated image.
Removing \textit{steps 100-400} significantly changes the parrot's appearance, but the new image retains the details and feather layering.
Removing \textit{steps 400-700} reduces the layering of the parrot's feathers.
Removing \textit{steps 700-1000} blurs the parrot's fur and the luster of the beak is gone, while it can retain a similar overall appearance to the original image.
Fig.~\ref{fig:motivation}(b) demonstrates the effect of adding an attribute in a specific stage.
In the \engordnumber{1} row, the sphere's appearance remains unchanged when injected the added concept ``yarn'' in \textit{steps 0-200}, but the background layout and colors are different, and adding it in \textit{steps 200-400} blurs the sphere's outline.
Injecting ``yarn'' in \textit{steps 400-600} and \textit{steps 600-800} leads to a more distinct texture. 
Adding ``yarn'' in \textit{steps 800-100} creates a woolen texture on the sphere and reduces its reflection.
The \engordnumber{2} row shows that the diffusion model is color-sensitive only at certain stages.
Fig.~\ref{fig:motivation}(c) shows the style removal results of impressionist Claude Monet and abstract painter Pablo Picasso.
We remove their names at different stages, i.e., using only ``a painting'' to guide the generation. 
Removing the style in \textit{steps 500-800} has little effect on the Picasso-guided painting, but the Monet-guided painting loses its brushstrokes. 
Conversely, removing \textit{steps 0-500} changes the content of the paintings guided by ``Monet'', but the style is maintained, while the image guided by ``Picasso'' loses its style.
We recommend zooming in to see experimental results of Monet's style.
In conclusion, the initial generation stages of the diffusion model tend to generate overall layout and color, the middle stages tend to generate structured appearances, and the final stages tend to generate detailed textures.

Based on the above observations, we can edit the results by changing the material, style, and content while keeping the layout unchanged by changing the prompts that act on different steps.
As shown in Fig.~\ref{fig:prompt_edit}, keeping the prompt ``lemon cake'' condition in the initial stages, the image can be edited into different appearances.
Prompt-to-prompt~\cite{hertz2022p2p} report the observation of similar effects and introduce a method that locks the corresponding attention maps.

\begin{figure}
\centering
\includegraphics[width=1\linewidth]{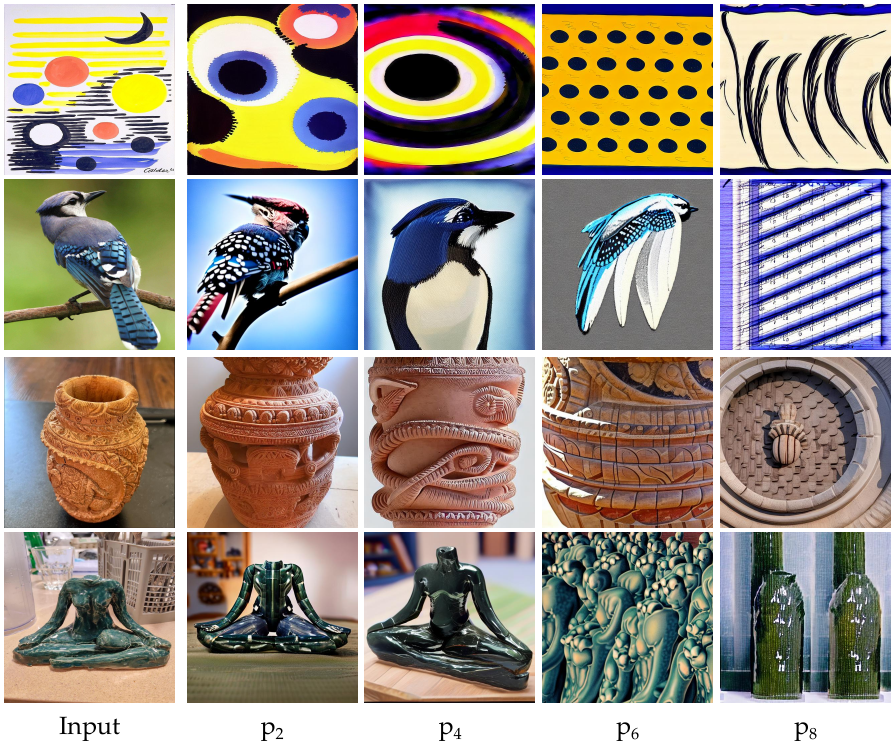}
\caption{The visualization results of token embeddings $p_i$ obtained by \sysname.
The results show that the initial generation step of the diffusion model is sensitive to structural information (e.g., bird's pose, pot's shape).
As the number of steps increases, the obtained $p_i$ gradually captures detailed information (e.g., the sideways head of the bird $\rightarrow$  bird's wing $\rightarrow$ the texture of the bird's feathers).
}
\label{fig:prompt_visualization}
\end{figure}

\subsection{Prompt Spectrum Space}
We use Stable Diffusion~\cite{latentdiffusion} as the generative backbone, which is built in the framework as Latent Diffusion Model (LDM)~\cite{latentdiffusion}.
LDM is a diffusion probability model that generates images by gradually denoising them.

Diffusion and denoising within an LDM typically take 1000 steps, and the text conditions the model step by step.
Previously, the process of the textual conditions acting on the diffusion model is regarded as a whole. 
In this work, we treat them as different procedures.
Specifically, we divide the 1000 steps of conditioning into ten stages on average.
Each stage corresponds to a unique textual condition.
The collection of textual conditions reside in the CLIP~\cite{clip} text-image space, their sizes are set to $n\times 1 \times 768$ ($n=10$ denotes the number of the stages).
This way of division is designed to keep a balance between efficiency and quality.

We refer to the expanded space as \textit{Prompt Spectrum Space}, denoted as $\STIspace$.
An illustration of how $\TIspace$ and $\STIspace$ interact with text and diffusion models is shown in Figs.~\ref{fig:intro}(a) and \ref{fig:intro}(b). 
Thus, $\STIspace$ is defined as:
\begin{equation}
    \begin{aligned}
        \STIspace=\{p_1,p_2,...,p_n\},
    \end{aligned}
\end{equation}
where $p_i$ represents the token embedding corresponding to the conditional prompt of the $i$th stage of the generation process.

\subsection{ProSpect}
\label{sec:STI}

We aim to extend TI~\cite{gal2022TI} to $\STIspace$ by extracting a \emph{set} of textual token embeddings from an input image.
To achieve this goal, we present \sysname, a method that maps an image to a collection of corresponding textual token embeddings.
The TI loss of LDM in $\TIspace$ space is formulated as:
\begin{equation}
\begin{aligned}
\loss_{TI}=\mathbb{E}_{z,t,p}\left[\left\|\epsilon-\epsilon_\theta\left(z_t, t, p_\theta\right)\right \|_2^2\right],
\end{aligned}
\end{equation}
where $p_\theta$ is a learnable vector denoting the token embedding and $z \sim E(x),\epsilon \sim \mathcal{N}(0, 1)$.
Similarly, the \sysname loss of LDM in $\STIspace$ space is formulated as:
\begin{equation}
\begin{aligned}
\loss_{PS}=\mathbb{E}_{z,t,p}\left[\left\|\epsilon-\epsilon_\theta\left(z_t, t, p_i\right)\right\| _2^2\right],
\label{eqn:loss_ps}
\end{aligned}
\end{equation}
where $p_i=P(t)$ is a learnable vector represents the token embedding of stage $i$, and $P=\left[p_1,p_2,...,p_n\right]$ is the set of textual token embeddings in $\STIspace$ space.

\begin{figure}
\centering
\includegraphics[width=\linewidth]{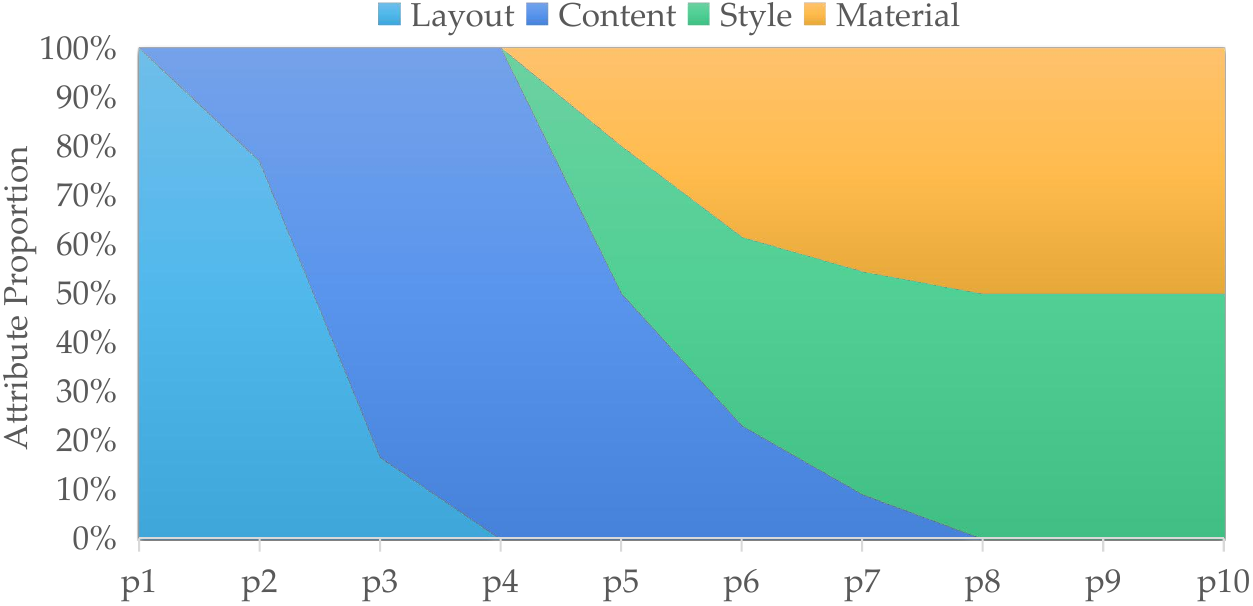}
\caption{
Statistical results of various attribute distributions at different prompts.
}
\label{fig:attribute_porpotion}
\end{figure}

\revision{
As shown in Fig.~\ref{fig:pipeline}(a), the token embedding is initialized to a frozen \(1\times768\) text embedding with a user input text (e.g., ``cup'') via the CLIP text encoder. It is then fed into a randomly initialized hypernetwork and finally creates a \(n\times1\times768\) embedding \(P=[p_1,p_2,..,p_n]\).
Only the hypernetwork is trainable and the final $p_i$ is obtained by optimizing based on Eqn.~(\ref{eqn:loss_ps}).
The training process typically requires 1000-3000 iterations.
Dropout is applied to prevent overfitting and the rate is set to 0.1.

Attribute control during inference is achieved by replacing the \(p_i\) representing different attributes with editing texts.
For instance, in Fig.~\ref{fig:pipeline}(b), content personalization involves maintaining the content-related \(p_3-p_{10}\) of image \textit{barn} as ``* in the jungle'' and replacing \(p_1-p_2\) with ``in the jungle'' (without ``*'' ).}

\section{Analysis of Prompt Spectrum Space}

\begin{table*}
\centering
\caption{CLIP-based evaluation results.
The best numbers are in \textbf{bold} and the second best results are \underline{underlined}.}
\resizebox{0.9\textwidth}{!}{
\begin{tabular}{c||c|ccc||c|ccc}
\toprule
Metric&\multicolumn{4}{c||}{Text Similarity$\uparrow$}&\multicolumn{4}{c}{Image Similarity$\uparrow$}\\
\hline
Method&Reference&\textbf{ProSpect}&DreamBooth&TI&Reference&\textbf{ProSpect}&DreamBooth&TI\\
\hline \hline
Average&0.2479&\textbf{0.3444}&\underline{0.3334}&0.3115&0.9128&\underline{0.7927}&\textbf{0.7987}&0.7274\\
\hline
Min&0.2168&\textbf{0.2869}&0.2279&\underline{0.2371}&0.8771&\textbf{0.6899}&\underline{0.6450}&0.4471\\
\hline
Max&0.2767&\textbf{0.3995}&0.3666&\underline{0.3820}&0.9541&\textbf{0.929}&\underline{0.8678}&0.8688\\
\hline
Negative Error&0.0311&0.0575&0.1055&0.0743&0.0357&0.1027&0.1537&0.2803\\
\hline
Positive Error&0.0288&0.0551&0.0331&0.0705&0.0412&0.1363&0.0691&0.1414\\
\bottomrule
\end{tabular}
}
\label{tab:clip_evaluation}
\end{table*}

\subsection{Visualization of Token Embeddings}

We visualize the token embedding $p_i$ obtained via \textit{ProSpect} by using it as the condition of the entire stage of the diffusion model, i.e., $p_{1:10}=p_i$.
Fig.~\ref{fig:prompt_visualization} shows the corresponding visual results of $p_i$ for four stages. 
It can be seen that the diffusion model acts different optimizations to token embeddings $p_i$ at different stages to reconstruct the given image.
The token embeddings that are conditioned on the initial stages are optimized to denote structure information, and then gradually represent detailed information as the generation steps increase.
For instance, $p_2$ tends to represent the layout or content, while $p_8$ tends to express the textures or brushstrokes.
The results indicate that different generation tendencies exist in different stages of the diffusion model.

\begin{figure}
\centering
\includegraphics[width=1\linewidth]{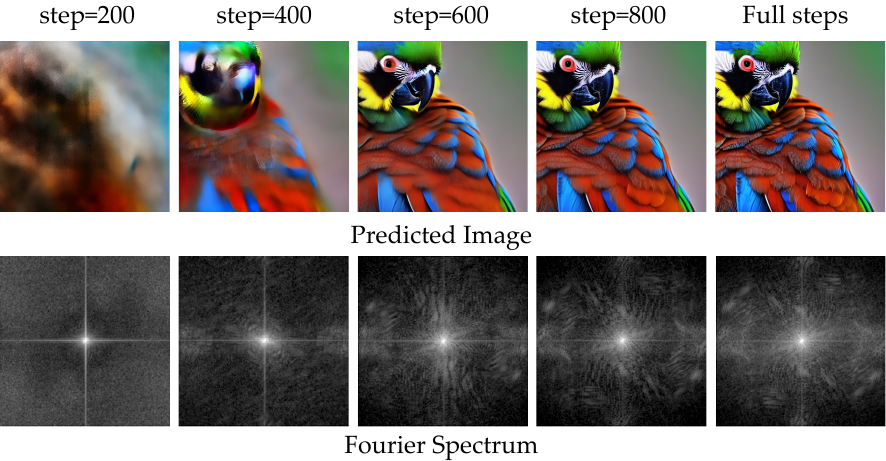}
\caption{
Analysis of images generated at different stages in the frequency domain.
The \engordnumber{1} row shows the predicted image obtained at different denoising steps with the text prompt ``\textit{a close-up photo of a parrot}''.
The \engordnumber{2} row showcases the Fourier spectrum of each predicted image.
As the denoising process progresses, the high-frequency information contained in the predicted image gradually increases.
We enhance the contrast of the Fourier spectrum for clarity.
}
\label{fig:fourier}
\end{figure}

\begin{figure*}
\centering
\includegraphics[width=1\linewidth]{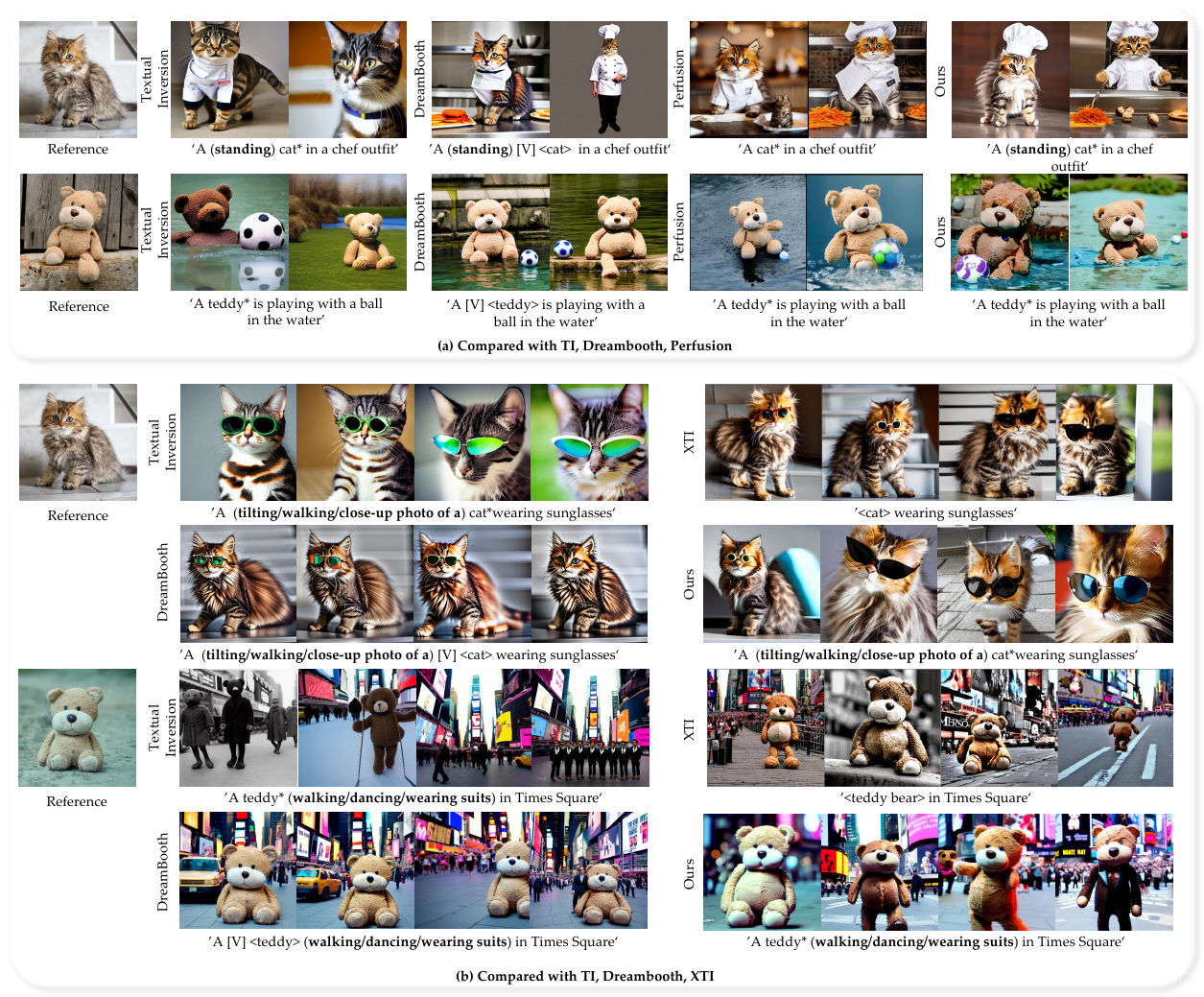}
\caption{Comparisons with state-of-the-art personalization methods including Textual Inversion (TI)~\cite{gal2022TI}, DreamBooth~\cite{ruiz2022dreambooth}, XTI~\cite{voynov2023pplus}, and Perfusion~\cite{Tewel:2023:perfusion}.
The \textbf{bold} words correspond to the additional concepts added to each image, (e.g. the \engordnumber{3} column in (a) shows the result of ``\textit{A standing cat in a chef outfit}'', the \engordnumber{6} column in (b) shows the result of ``\textit{A tilting cat wearing sunglasses}'').
XTI and Perfusion are the latest published methods and the model have not been released yet. 
The resulting images of XTI and Perfusion are borrowed from their paper, so the results of adding concepts are not shown.
Our method can faithfully convey the appearance and material of the reference image with better controllability and diversity.
}
\label{fig:inversion_sota}
\end{figure*}

\revision{
\subsection{Visualization of Attribute Distribution}

To evaluate the attribute distribution, we provide 30 pairs of \textit{attribute, object} combinations (e.g., ``origami, cake''), including 10 pairs for material, style, and layout, respectively.
The \textit{object} remains unchanged while we record the impact of adding \textit{attribute} at different $p_i$. Additionally, we select 10 new \textit{objects} to replace the original \textit{object} at different $p_i$ and record the impact of replacement on the content.
The results are shown in Fig.~\ref{fig:attribute_porpotion}.
Notably, adding attributes or replacing content at a single $p_i$ may not significantly change the output image.
To ensure a faithful evaluation, we gradually increase the intensity of the change until other attributes are affected. 
}

\subsection{Explanations}

The experimental results demonstrate that the diffusion model generates images in the order of \textit{layout $\rightarrow$ content $\rightarrow$material/style}.
A similar phenomenon has been observed in convolutional networks.
\citet{voynov2023pplus} noted that the U-Net structure of the diffusion model has similar properties, with the shallow layer tending to generate texture and color and the deep layer generating semantic information.
It is important to note that the deep receptive field size of U-Net is larger than the shallow receptive field size, making the hierarchical attribute distribution easy to comprehend.
However, this size difference dose \textit{not} exist between steps of the diffusion model, since the latent size is uniform across different stages.

The Fourier transform is a classic transformation widely used in digital image processing.
It transforms a signal from the time domain into the frequency domain, facilitating the identification of subtle features and the processing of challenging components.

\begin{figure}
\centering
\includegraphics[width=1\linewidth]{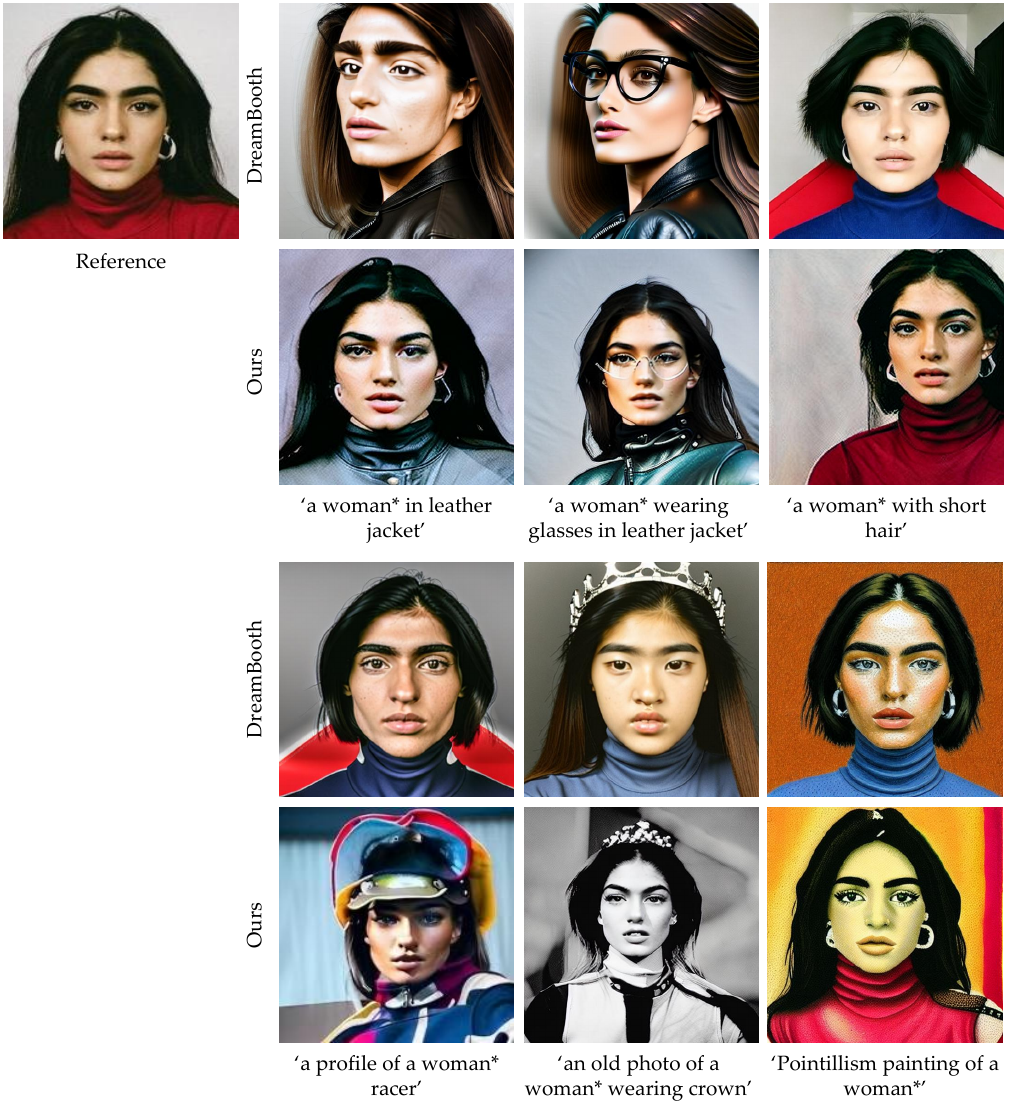}
\caption{Comparision with DreamBooth~\cite{ruiz2022dreambooth} on personalized one-shot portrait generation.
Our inversion based method can better preserve the character identity in the input image.
}
\label{fig:inversion_woman}
\end{figure}

Fig.~\ref{fig:fourier} shows the Fourier spectrum of the diffusion process.
As the number of steps in the denoising process increases, the high-frequency information contained in the image predicted by the diffusion model gradually increases.
This indicates that the model tends to generate structural information at the beginning of the denoising process, with details gradually increasing as the steps increase.
This phenomenon explains the generation order of the diffusion model, which is caused by the signal frequency of the corresponding attribute from low to high.

\begin{figure*}
\centering
\includegraphics[width=1\linewidth]{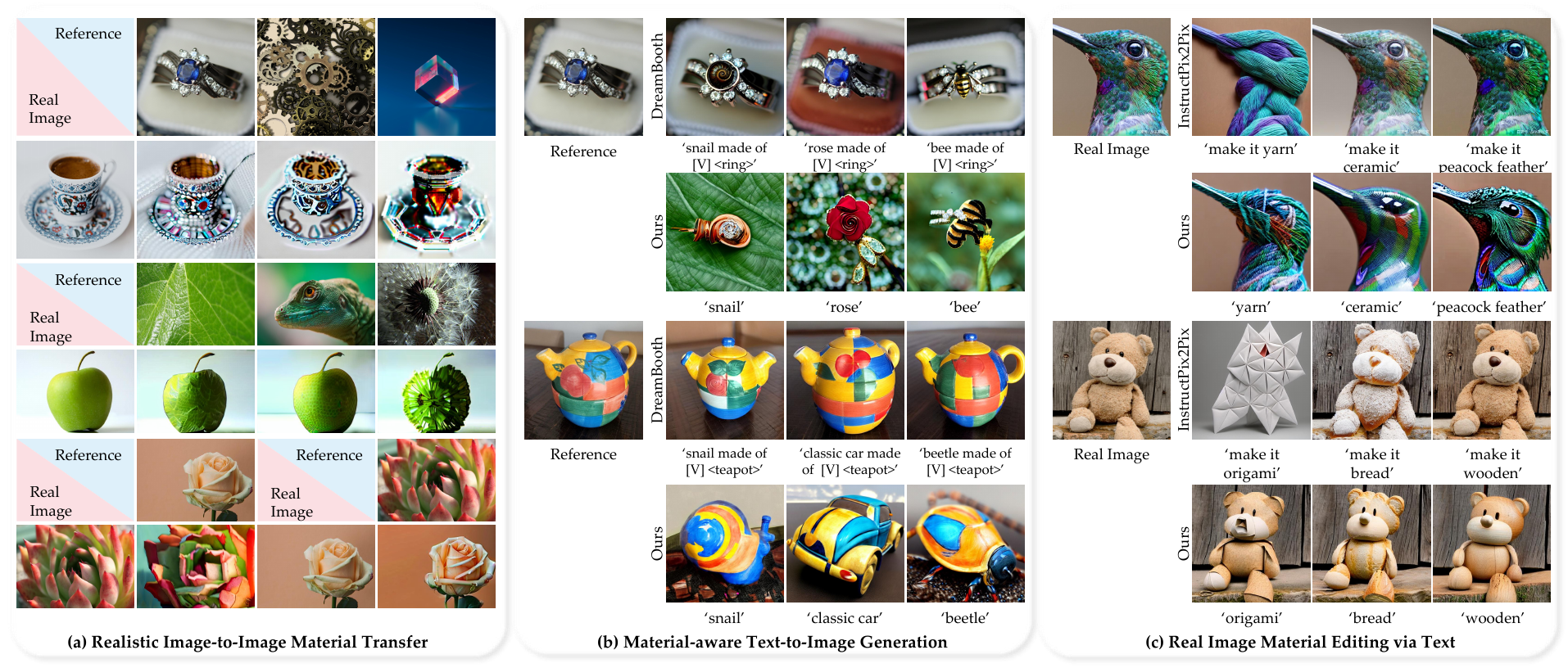}
\caption{Material-aware image generation results.
We compare \sysname with a personalization approach DreamBooth~\cite{ruiz2022dreambooth} and an image editing approach InstructPix2Pix~\cite{brooks2022instructpix2pix}.
Our method shows better fidelity and editability.
}
\label{fig:texture_result}
\end{figure*}

\section{Experiments}
\label{sec:experiments}

We demonstrate that \sysname outperforms state-of-the-art text-to-image personalization baselines in both fidelity and editability by conducting both qualitative and quantitative evaluations.
Moreover, we apply \sysname to diverse applications of material transfer, style transfer, and layout transfer (as shown in Sec.~\ref{sec:application}), and perform qualitative comparisons with related methods.

\paragraph{Methods for comparison}
\revision{
We optimize \textbf{(1) Textual Inversion (TI)}~\cite{gal2022TI} with 5000 iterations and \textbf{(2) InST}~\cite{zhang2023inst} with 1000 iterations on Stable Diffusion 1.4~\cite{latentdiffusion}, both as recommended by the authors.
We train \textbf{(3) DreamBooth}~\cite{ruiz2022dreambooth} for 400 steps.
The resulting images of \textbf{(4) Perfusion}~\cite{Tewel:2023:perfusion} and \textbf{(5) XTI}~\cite{voynov2023pplus} are borrowed from their papers.
We use the official pre-trained models of \textbf{(6) InstructPix2Pix}~\cite{brooks2022instructpix2pix}, \textbf{(7) JoJoGAN}~\cite{chong2022jojogan}, \textbf{(8) CAST}~\cite{zhang2022cast}, and \textbf{(9) StyTr$^2$}~\cite{deng2022stytr2}.
}
\revision{
\paragraph{Test dataset}
For fair comparison, we use nine concepts from previous papers, including cat, teddy bear, cat statue, pot, sculpture, colorful teapot, red teapot, elephant, clock, and three concepts of faces.
For each concept, we use three easy prompts (changing background) and three difficult prompts (changing pose/clothes/views/etc.).
Each image-prompt pair is used to generate four results.
In total, we obtain 288 images for each method.
}

\paragraph{Implementation details}
In all of our experiments, we use Stable Diffusion 1.4~\cite{latentdiffusion} with the default hyperparameters and set a base learning rate of 0.001. 
We employ a DDIM sampler with diffusion steps $T = 50$ and guidance scale $w = 7.5$. 
We use a frozen CLIP model in Stable Diffusion as the text encoder network.
The texts are tokenized into start-token, end-token, and 75 non-text padding tokens.
The training process on each image takes approximately 20 minutes using an NVIDIA GeForce RTX3090 with a batch size of 1, significantly less than the more than 90 minutes required for TI.
The synthesis process takes about three seconds, depending on the number of diffusion steps taken.

\subsection{Quantitative Evaluation}

We use two metrics to conduct quantitative evaluations.
Specifically, we compute the pair-wise CLIP cosine similarity between the reference images and the generated images as \emph{image similarity} to evaluate content fidelity.
In addition, we use the CLIP similarity between all generated images and their textual conditions as \emph{text similarity} to evaluate the editability.

Table~\ref{tab:clip_evaluation} shows the corresponding quantitative evaluation results of our method and two baseline methods.
The Reference column of text similarity calculates the cosine similarity between the reference image and the various text condition, which can be regarded as the lower bound score.
The Reference column of image similarity calculates the cosine similarity between the image contains the same object and the reference image, which can be regarded as the groundtruth score.
TI~\cite{gal2022TI} fails to preserve object appearance, while DreamBooth tends to overfit the reference image. 
Though a higher fidelity score it gets,  the editability is not satisfactory.
Our method achieves a better balance of object fidelity and editability without fine-tuning the model.

\begin{figure*}
\centering
\includegraphics[width=1\linewidth]{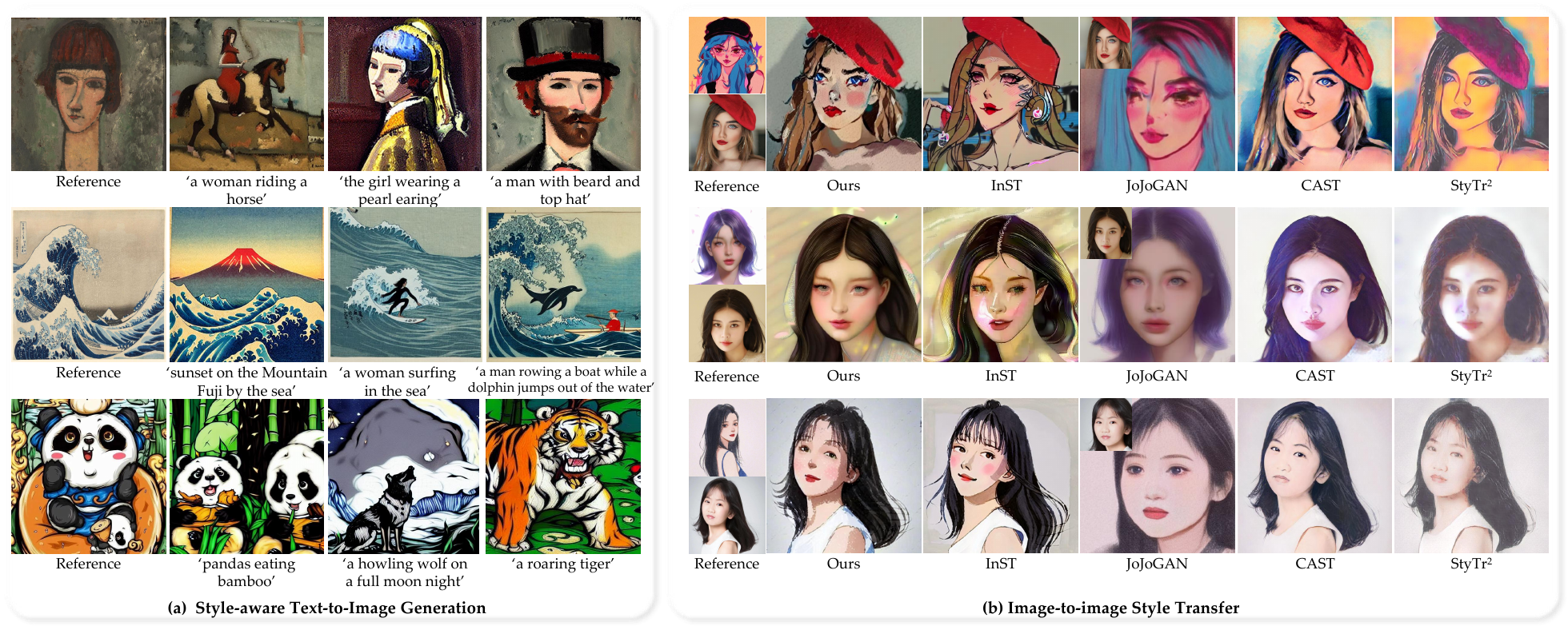}
\caption{
Style-aware image generation results.
\revision{We compare \textit{ProSpect} with state-of-the-art style transfer methods, including InST~\cite{zhang2023inst}, JoJoGAN~\cite{chong2022jojogan}, CAST~\cite{zhang2022cast}, and StyTr$^2$~\cite{deng2022stytr2}.
Our method better preserves the identity information of the content image than the diffusion-based method InST while generating better brush strokes than other GAN-based and encoder-based methods.
Style image credits (the \engordnumber{1} and \engordnumber{2} rows in (a)): \{Amedeo Modigliani, Katsushika Hokusai\}/The Art Institute of Chicago (CC0)~\cite{AIC}. }
}
\label{fig:style_result}
\end{figure*}
\begin{figure*}
\centering
\includegraphics[width=\linewidth]{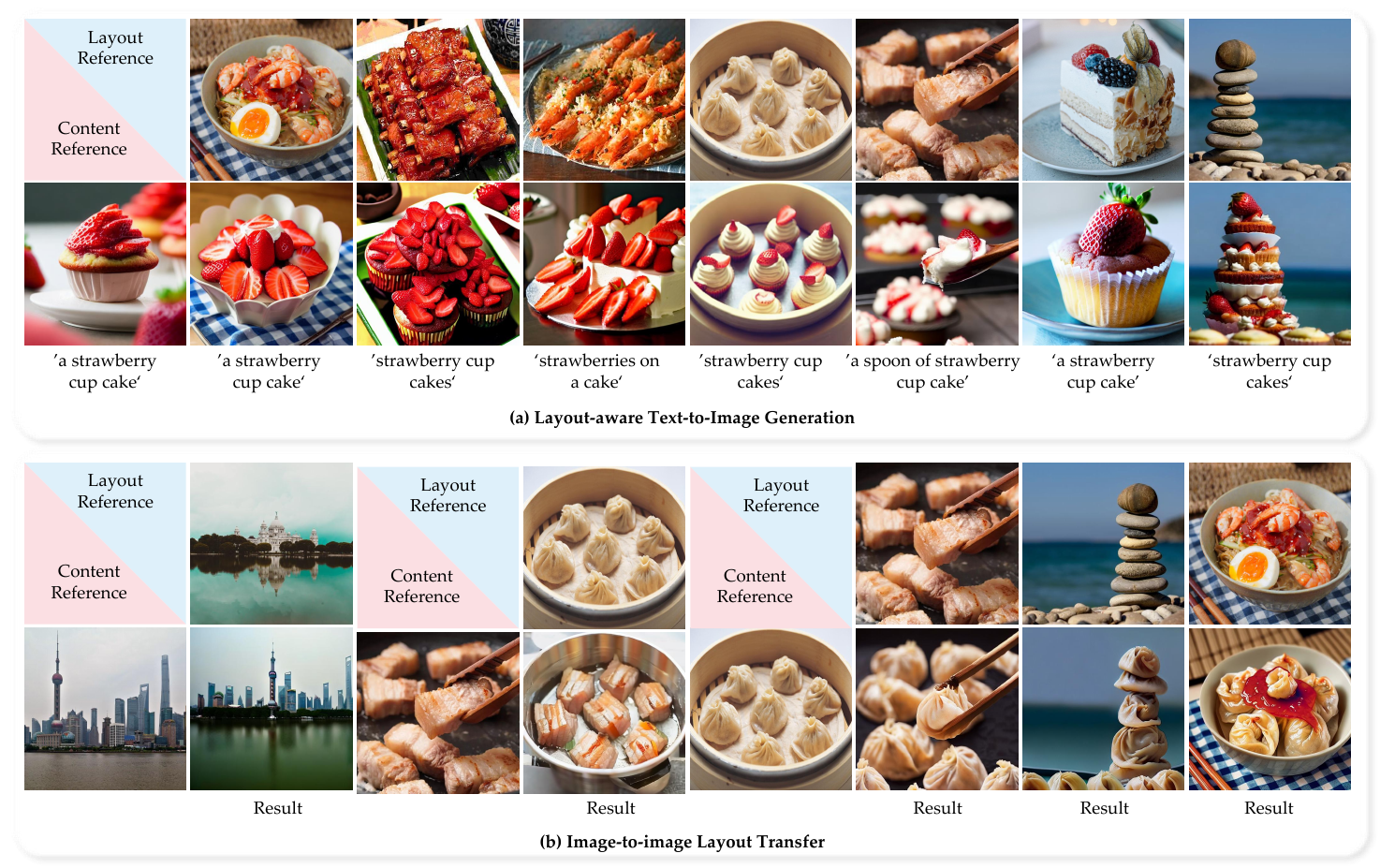}
\caption{
Layout-aware image generation results.
\sysname can generate an image with the same layout of an layout reference image by using a text prompt or a content reference image.
}
\label{fig:layout_result}
\end{figure*}

\begin{figure}
\centering
\includegraphics[width=1\linewidth]{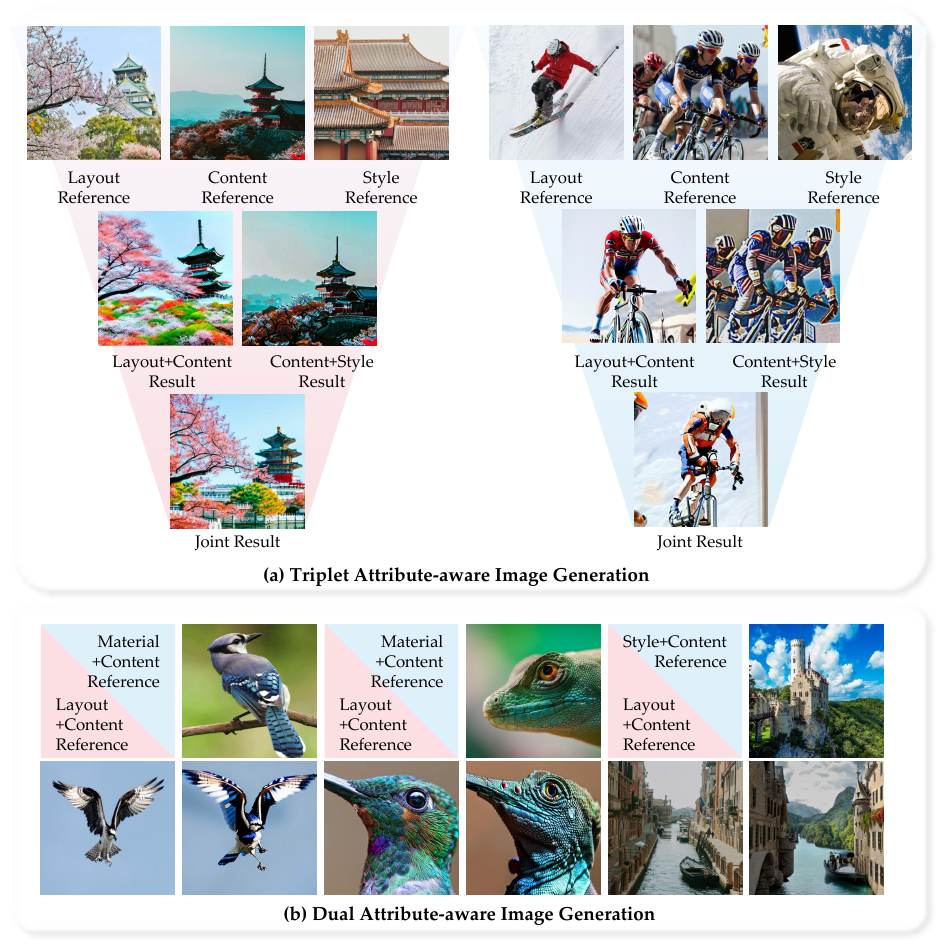}
\caption{\revision{Results of multi-attribute-aware image generation with \sysname.
(a) Each reference offers one kind of visual attribute, and we combine them progressively to generate joint results by mixing the triplet references.
(b) Each reference indicates two kinds of visual attributes, and we mix two references by taking the material/layout/style attribute from individual references and scaling the range of content conditions. }
}
\label{fig:joint_result}
\end{figure}


\subsection{Qualitative Evaluation}

As shown in Fig.~\ref{fig:inversion_sota}, we compare our method with four SOTA personalization methods, \ie, TI~\cite{gal2022TI}, DreamBooth~\cite{ruiz2022dreambooth}, XTI~\cite{voynov2023pplus}, and Perfusion~\cite{Tewel:2023:perfusion}.
We use concepts from previous papers for fair comparison and unbiased evaluation.
We add additional texts shown in bold to each set of images to demonstrate the flexibility of our method.

DreamBooth can well depict the conceptual appearance in the reference image, but tends to overfit to the reference image, resulting in a lack of editability.
As shown in the results of {``a (standing) cat in a chef outfit''} in the second row, TI fails to maintain the object's appearance and generates normal cats.
DreamBooth can generate a standing cat, but the background is blurred, and the cat's paw is confused with the human hand.
Our results can generate a standing cat with a kitchen as the background and maintain the details of the cat's paws.
The results of {``a (tilting/walking/close-up photo of a) cat wearing sunglasses''} show that DreamBooth can generate a cat with sunglasses, but cannot change the cat's posture or zoom-in/zoom-out.
Our method, shown in the third row, can generate high-fidelity concepts while maintaining diversity and flexibility.
\sysname not only puts sunglasses on the cat but also allows it to show its walking posture and close-up details.
In the results of {``a teddy is playing with a ball in the water''},  Perfusion and DreamBooth can generate teddy bear, ball, and water, but they are not interacting with each other.
Our method can show the posture of the teddy bear touching and throwing the ball, and the teddy bear can float on the water or half-submerge in the water.
In the results of {``a teddy (walking/dancing/wearing suits) in Times Square''}, XTI cannot accurately maintain the appearance of the teddy bear, and DreamBooth cannot change the posture of the teddy bear.
Our method can reproduce the appearance of a teddy bear while walking, dancing, and wearing a suit, always in the background of Times Square.

Our method is also capable of personalized one-shot portrait generation.
Fig.~\ref{fig:inversion_woman} shows the comparison results between our method and DreamBooth~\cite{ruiz2022dreambooth}.
Our method can manipulate attributes such clothing, hairstyle and artistic styles of the input portrait while preserving the identity.

\subsection{User Study}	

We evaluate our method in attributes-aware image generation, alongside three SOTA personalization methods, \ie, TI~\cite{gal2022TI}, DreamBooth~\cite{ruiz2022dreambooth}, and InST~\cite{zhang2023inst}.
A total of 66 participants took part in the survey, including 42 researchers in computer graphics or computer vision (CGCV), 24 university students (others).
The user study is divided into three parts, including personalized objects, material guidance, and style guidance. 

\paragraph{User Study \uppercase\expandafter{\romannumeral1}}
In the content-aware image generation survey, TI and DreamBooth are used as the baseline methods.
\revision{
The same 12 concepts in quantitative evaluation, each with two different prompts are used.}
The objective of the personalization task, which is to generate a new image with the same concept as the reference image while also matching the provided text condition, is introduced to the participants.
For each question, the participants are shown a reference image and a text condition (e.g., ``{a photo of the same cat wearing sunglasses}'') and are asked to choose the option that best matches the task objective from three randomly ordered options, each corresponding to a method.
$\sysname$ receives $51.97\%$ \reviseagain{(CGCV $52.14\%$, Others $51.67\%$)} of the preferences, while TI acquires $10.30\%$ \reviseagain{(CGCV $9.76\%$, Others $11.25\%$)}, and DreamBooth obtains $37.72\%$ \reviseagain{(CGCV $38.09\%$, Others $37.08\%$)}.
Thus, \sysname exhibits better performance in human preference when compared to the two baseline methods.

\paragraph{User Study \uppercase\expandafter{\romannumeral2}}
In the material-aware image generation survey, DreamBooth is used as the baseline method, and the participants are introduced that the objective of the task is to generate a new image composed of materials from the reference image while matching the provided text conditions. 
\revision{
Eight material references with three results each are used.}
For each question, the participants are shown reference images and corresponding text conditions (e.g., ``{a snail made of the material in this image}'') and are asked to select one of two options that best matches the task objective.
$\sysname$ receives $66.36\%$'s preference \reviseagain{(CGCV $68.57\%$, Others $62.50\%$)} and DreamBooth obtains $33.64\%$ \reviseagain{(CGCV $31.42\%$, Others $37.50\%$)}.

\paragraph{User Study \uppercase\expandafter{\romannumeral3}}
The SOTA style transfer method InST~\cite{zhang2023inst} is the baseline method in the style-aware image generation survey.
\revision{
Eight style references with one style transfer result and one T2I result each are used.}
We evaluate both the style-guided text-to-image generation task and the style transfer task.
The participants are introduced that the objective of the task is to generate a new image consistent with the style of the reference artistic image while also being consistent with the content of the provided textual condition/content image.
For each question, the participants are presented with either a style image and a corresponding text condition (e.g., ``a painting of Einstein drawn in the style of the reference image'') or a pair of style and content images, and are asked to select one of two options that best matches the task objective.
$\sysname$ outperforms InST by receiving $61.67\%$ \reviseagain{(CGCV $61.19\%$, Others $62.50\%$)} the preference of compared with InST's $38.33\%$ \reviseagain{(CGCV $38.80\%$, Others $37.50\%$)}.

\subsection{Applications}
\label{sec:application}

In this section, we demonstrate the effectiveness of our approach in various attribute-aware image generation tasks, including material-aware image generation, style-aware image generation, as well as layout-aware image generation.

\paragraph{Material-aware image generation}

Our approach is well-suited for material-aware image generation tasks, including material transfer between images, image material-guided text-to-image generation, and image material editing with text.
Results shown in Fig.~\ref{fig:texture_result} demonstrate the high visual quality and flexibility of our method.
Fig.~\ref{fig:texture_result}(a) shows the results of material transfer, where our method can transfer materials between semantically unrelated objects (e.g., gears and teacups, apples, and dandelions). 
Fig.~\ref{fig:texture_result}(b) shows the material-guided text-to-image generation using a reference image, which we compare with a state-of-the-art personalization method DreamBooth~\cite{ruiz2022dreambooth}.
DreamBooth requires both prompt learning and model fine-tuning, making it prone to overfitting on specific images and lacking flexibility with single-image input.
Our method, however, can guide image generation using references with unrelated materials (e.g., rings and snails, teapot, and beetle), demonstrating superior editability.
Fig.~\ref{fig:texture_result}(c) shows the results of modifying an image's material with natural language.
We compare our method with a state-of-the-art image editing method InstructPix2Pix~\cite{brooks2022instructpix2pix}, which works on semantically related images (e.g., hummingbird to peacock feather) but fails on semantically unrelated modifications (e.g., teddy to origami).
Unlike InstructPix2Pix, our method can edit images into completely unrelated materials while retaining their overall appearance and background.

\paragraph{Style-aware image generation}
Our method is also effective for generating artistic images.
The material in a realistic image reflects high-frequency information, while strokes and shapes reflect the same in an artistic image.
Using a similar approach to material transfer, we can perform style transfer and style-guided text-to-image generation.
Fig.~\ref{fig:style_result}(a) shows the results of style-guided text-to-image generation, where our method learns the style from a single artistic image and generates new images that are semantically different (e.g., ``{an astronaut landing on a planet}'') or more vivid in content (e.g., ``{a man rowing a boat while a dolphin jumps out of the water}''), while accurately reproducing the reference image's style.
Fig.~\ref{fig:style_result}(b) shows the results of style transfer, comparing it with the state-of-the-art diffusion-based style transfer method InST~\cite{zhang2023inst}, the GAN-based method JoJoGAN~\cite{chong2022jojogan}, encoder-decoder-based method CAST~\cite{zhang2022cast}, and ViT-based method StyTr$^2$~\cite{deng2022stytr2}.
Since InST considers the overall appearance of an image as a condition and lacks disentanglement of style and content, the generated image often lacks identity. 
JoJoGAN needs to align the face key points of the content image and style image, so some special styles may cause artifacts and distortions (as shown in the \engordnumber{1} row), and the generated images may have content in-consistency (as shown in the \engordnumber{2} row). 
CAST and StyTr$^2$ fail to transfer the shape changes and large brushstrokes.
Our method produces more realistic strokes (e.g., the hair in \engordnumber{1} and \engordnumber{3} rows), fewer artifacts (e.g. the \engordnumber{2} row), and better-maintained identity.

\begin{figure}
\centering
\includegraphics[width=1\linewidth]{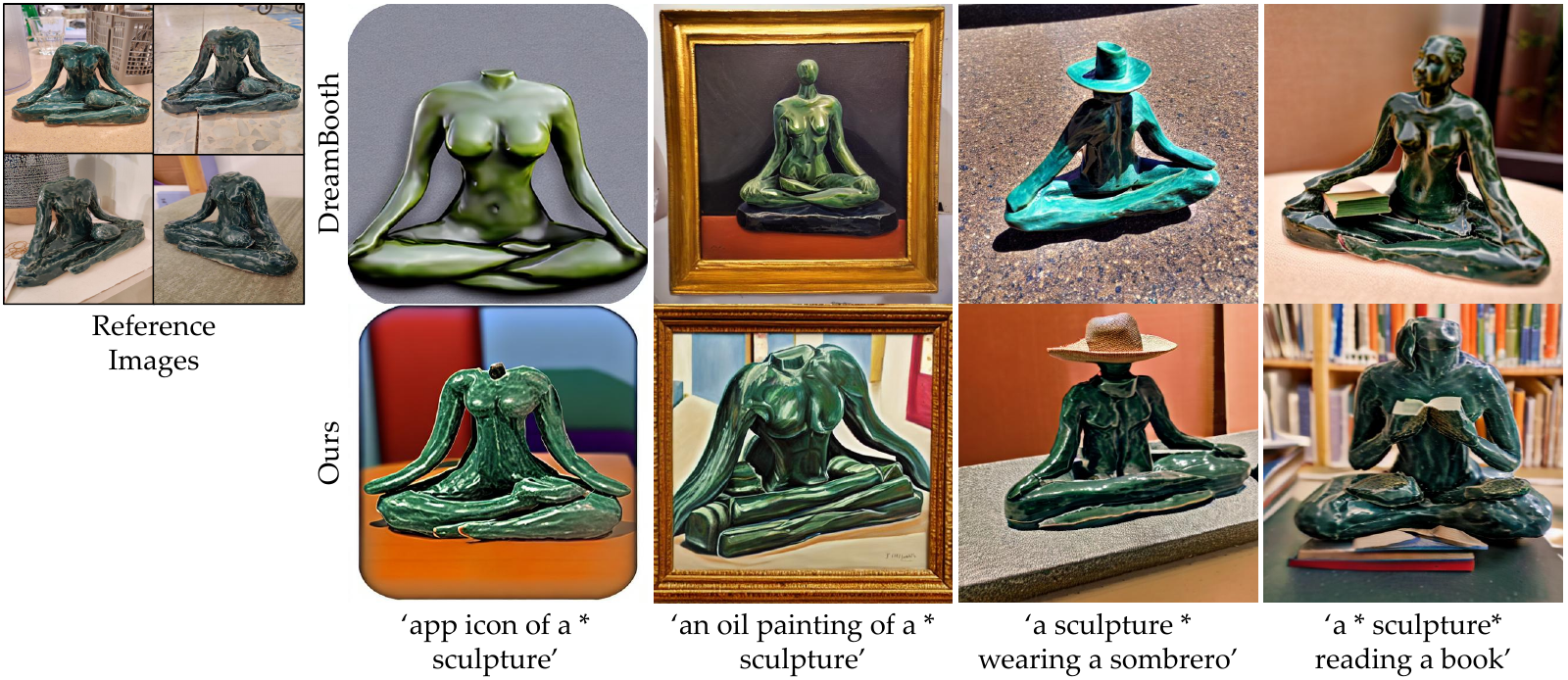}
\caption{\revision{Comparison of results by training with a small number of images.}
}
\label{fig:group_inversion}
\end{figure}
\begin{figure}
\centering
\includegraphics[width=1\linewidth]{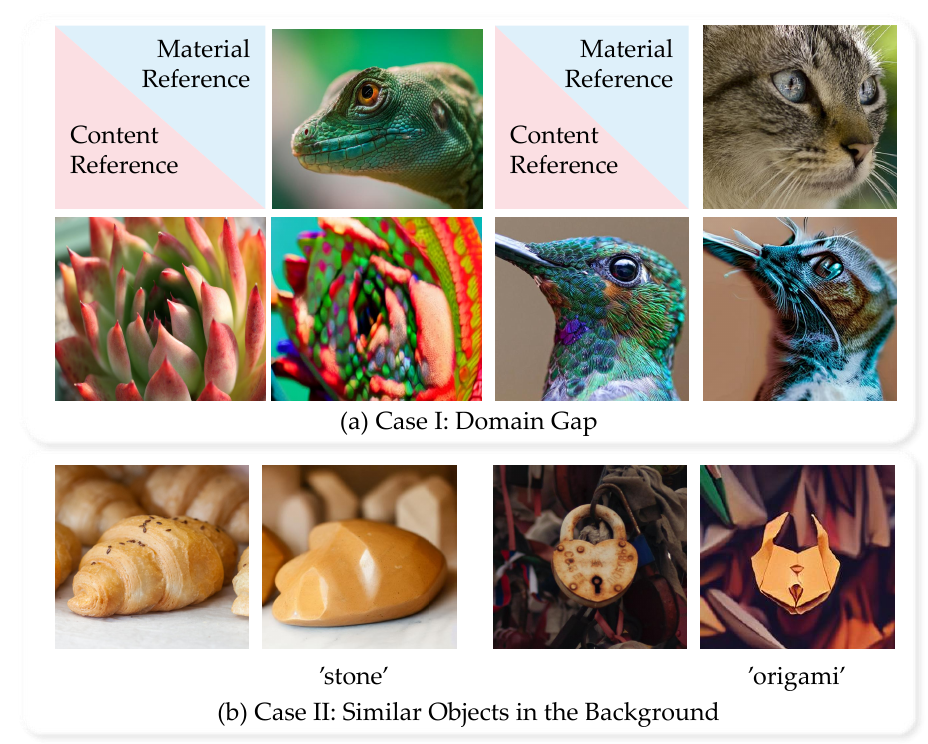}
\caption{Examples of failure cases.
(a) Results of transferring materials between images with large domain gaps.
(b) When the image background is composed of similar objects sharing the same frequency information, attribute editing may be applied to the entire image.
}
\label{fig:limitation}
\end{figure}

\paragraph{Layout-aware image generation}
Layout is a core element of photography that determines the quality of a photo.
The low-frequency information of an image reflects its layout.
By learning this information, our method can use the layout of a single given image to guide text-to-image generation and transfer the layout of an image to another image.
Fig.~\ref{fig:layout_result}(a) shows the results of layout-guided text-to-image generation, where our method learns complex composition (e.g., ``{a spoon of strawberry cupcake}'') and guides the generation of semantically unrelated content (e.g., strawberry cupcake and rock) from a reference image.
Fig.~\ref{fig:layout_result}(b) displays the results of layout transfer for landscape and still-life images.
Our method can transfer the ``centering'' and ``reflection'' features of a photo to another landscape image (see the second column in Fig.~\ref{fig:layout_result}(b)) and transfer complex object layouts to another still-life image.

\paragraph{Multi-attribute-aware image generation}
In Fig.~\ref{fig:joint_result}, we combine attributes from multiple images to guide the generation process. 
In Fig.~\ref{fig:joint_result}(a), the layout, content, and style are guided by three reference images. 
Results for a landscape example are shown in the left pink pyramid. The first row displays reference images, the second row displays results using dual-attribute guidance, and the bottom row shows the result using triple-attribute guidance. The bottom result maintains the relative position of the flowers and architecture in the layout image, has the three-floor building structure from the content reference, and replicates the appearance of Chinese architecture from the style reference. 
In the right blue pyramid, we show results for a portrait example. The result is guided by the layout of a single person in the middle, the content of a cyclist, and the style of an astronaut.
Fig.~\ref{fig:joint_result}(b) shows a different setting by mixing multiple attributes from one image.

\paragraph{Few-shot image generation}
$\sysname$ is designed to accept a single image as input, but it can also work on a set of images, similar to DreamBooth~\cite{ruiz2022dreambooth}.
As shown in Fig.~\ref{fig:group_inversion}, \sysname can produce results with improved fidelity and diversity compared to prior approaches when applied to four sculpture images. 
In addition, $\sysname$ can also be applied to model fine-tuning methods.

\subsection{Limitations}
\label{sec:discussion}

First, although $\sysname$ is faster than TI~\cite{gal2022TI}, it is still not as fast as some encoder-based methods~\cite{gal2023encoder}, given that each iteration of optimization is calculated on a random step and  $\sysname$ learns several token embeddings at different steps.
Second, as shown in Fig.~\ref{fig:limitation}(a),  $\sysname$ can achieve attribute disentanglement, but the attribute transfer between images with large domain gap may not be visually aesthetic.
Finally, Fig.~\ref{fig:limitation}(b) shows the cases of dealing with images in which the background is composed of similar objects.
Since the objects of the same category are of similar scales, sometimes the attribute modification may act on the background objects undesirably.

\section{Conclusion and Future Work}
\label{sec:conclusion}

In this paper, we delve into the image generation process of the diffusion model from the perspective of steps.
We propose an expanded textual conditioning space, denoted by $\STIspace$, for diffusion models.
Our experiments demonstrate that $\STIspace$ has better disentanglement and controllability, allowing for generating images from different granularities.
To further enable images to be represented in $\STIspace$, we propose $\sysname$, which inverts the text conditions of the diffusion model step by step.
$\sysname$ provides more fidelity and editable image representations, paving the way for attributes-aware image generation.
Using $\sysname$, material/style/content/layout-related transfer and editing tasks can be performed.
Our evaluations and experimental results demonstrate that $\sysname$ offers superior fidelity, expressiveness, and controllability for diverse image generation tasks.
In the future, we plan to further develop and improve methods for attribute disentanglement, such as making a more detailed attribute division and recombination methods as well as studying the mutual impact of different textual conditions.

\begin{acks}
This work was supported in part by National Key R\&D Program of China under no. 2020AAA0106200, by National Natural Science Foundation of China under nos. 61832016, 62102162, and U20B2070, in part by Beijing Natural Science Foundation under no. L221013, in part by the National Science and Technology Council under no. 111-2221-E-006-112-MY3, Taiwan, and in part by the Deutsche Forschungsgemeinschaft (DFG) under no. 413891298.
\end{acks}

\bibliographystyle{ACM-Reference-Format}
\balance
\bibliography{ProSpect}


\begin{thebibliography}{128}


\ifx \showCODEN    \undefined \def \showCODEN     #1{\unskip}     \fi
\ifx \showDOI      \undefined \def \showDOI       #1{#1}\fi
\ifx \showISBNx    \undefined \def \showISBNx     #1{\unskip}     \fi
\ifx \showISBNxiii \undefined \def \showISBNxiii  #1{\unskip}     \fi
\ifx \showISSN     \undefined \def \showISSN      #1{\unskip}     \fi
\ifx \showLCCN     \undefined \def \showLCCN      #1{\unskip}     \fi
\ifx \shownote     \undefined \def \shownote      #1{#1}          \fi
\ifx \showarticletitle \undefined \def \showarticletitle #1{#1}   \fi
\ifx \showURL      \undefined \def \showURL       {\relax}        \fi
\providecommand\bibfield[2]{#2}
\providecommand\bibinfo[2]{#2}
\providecommand\natexlab[1]{#1}
\providecommand\showeprint[2][]{arXiv:#2}

\bibitem[{$\!\!$}(1984)]%
        {1984:1040142}
 \bibinfo{year}{1984}\natexlab{}.
\newblock \bibinfo{journal}{\emph{SIGCOMM Comput. Commun. Rev.}} \bibinfo{volume}{13-14}, \bibinfo{number}{5-1} (\bibinfo{year}{1984}).
\newblock
\showISSN{0146-4833}


\bibitem[Cze(2008)]%
        {Czerwinski:2008:1358628}
 \bibinfo{year}{2008}\natexlab{}.
\newblock \bibinfo{booktitle}{\emph{CHI '08: CHI '08 extended abstracts on Human factors in computing systems}} (Florence, Italy). \bibinfo{publisher}{ACM}, \bibinfo{address}{New York, NY, USA}.
\newblock
\showISBNx{978-1-60558-012-X}
\newblock
\shownote{General Chair-Czerwinski, Mary and General Chair-Lund, Arnie and Program Chair-Tan, Desney}.


\bibitem[Ablamowicz and Fauser(2007)]%
        {Ablamowicz07}
\bibfield{author}{\bibinfo{person}{Rafal Ablamowicz} {and} \bibinfo{person}{Bertfried Fauser}.} \bibinfo{year}{2007}\natexlab{}.
\newblock \bibinfo{booktitle}{\emph{CLIFFORD: a Maple 11 Package for Clifford Algebra Computations, version 11}}.
\newblock
\urldef\tempurl%
\url{http://math.tntech.edu/rafal/cliff11/index.html}
\showURL{%
Retrieved February 28, 2008 from \tempurl}


\bibitem[Abril and Plant(2007)]%
        {Abril07}
\bibfield{author}{\bibinfo{person}{Patricia~S. Abril} {and} \bibinfo{person}{Robert Plant}.} \bibinfo{year}{2007}\natexlab{}.
\newblock \showarticletitle{The patent holder's dilemma: Buy, sell, or troll?}
\newblock \bibinfo{journal}{\emph{Commun. ACM}} \bibinfo{volume}{50}, \bibinfo{number}{1} (\bibinfo{year}{2007}), \bibinfo{pages}{36--44}.
\newblock
\urldef\tempurl%
\url{https://doi.org/10.1145/1188913.1188915}
\showDOI{\tempurl}


\bibitem[Andler(1979)]%
        {Andler79}
\bibfield{author}{\bibinfo{person}{Sten Andler}.} \bibinfo{year}{1979}\natexlab{}.
\newblock \showarticletitle{Predicate Path expressions}. In \bibinfo{booktitle}{\emph{Proceedings of the 6th. ACM SIGACT-SIGPLAN symposium on Principles of Programming Languages}} \emph{(\bibinfo{series}{POPL '79})}. \bibinfo{publisher}{ACM Press}, \bibinfo{address}{New York, NY}, \bibinfo{pages}{226--236}.
\newblock
\urldef\tempurl%
\url{https://doi.org/10.1145/567752.567774}
\showDOI{\tempurl}


\bibitem[Anisi(2003)]%
        {anisi03}
\bibfield{author}{\bibinfo{person}{David~A. Anisi}.} \bibinfo{year}{2003}\natexlab{}.
\newblock \emph{\bibinfo{title}{Optimal Motion Control of a Ground Vehicle}}.
\newblock \bibinfo{thesistype}{Master's\ thesis}. \bibinfo{school}{Royal Institute of Technology (KTH), Stockholm, Sweden}.
\newblock


\bibitem[{Art Institute of Chicago}(2023)]%
        {AIC}
\bibfield{author}{\bibinfo{person}{{Art Institute of Chicago}}.} \bibinfo{year}{2023}\natexlab{}.
\newblock
\newblock
\urldef\tempurl%
\url{https://www.artic.edu/}
\showURL{%
\tempurl}
\newblock
\shownote{Last accessed on 2023-09-12}.


\bibitem[Avrahami et~al\mbox{.}(2022)]%
        {avrahami2022blended}
\bibfield{author}{\bibinfo{person}{Omri Avrahami}, \bibinfo{person}{Dani Lischinski}, {and} \bibinfo{person}{Ohad Fried}.} \bibinfo{year}{2022}\natexlab{}.
\newblock \showarticletitle{Blended Diffusion for Text-Driven Editing of Natural Images}. In \bibinfo{booktitle}{\emph{IEEE/CVF Conference on Computer Vision and Pattern Recognition (CVPR)}}. \bibinfo{pages}{18208--18218}.
\newblock


\bibitem[Balaji et~al\mbox{.}(2022)]%
        {balaji2022ediffi}
\bibfield{author}{\bibinfo{person}{Yogesh Balaji}, \bibinfo{person}{Seungjun Nah}, \bibinfo{person}{Xun Huang}, \bibinfo{person}{Arash Vahdat}, \bibinfo{person}{Jiaming Song}, \bibinfo{person}{Karsten Kreis}, \bibinfo{person}{Miika Aittala}, \bibinfo{person}{Timo Aila}, \bibinfo{person}{Samuli Laine}, \bibinfo{person}{Bryan Catanzaro}, \bibinfo{person}{Tero Karras}, {and} \bibinfo{person}{Ming-Yu Liu}.} \bibinfo{year}{2022}\natexlab{}.
\newblock \showarticletitle{{eDiff-I:} Text-to-Image Diffusion Models with an Ensemble of Expert Denoisers}.
\newblock \bibinfo{journal}{\emph{arXiv preprint arXiv:2211.01324}} (\bibinfo{year}{2022}).
\newblock


\bibitem[Bau et~al\mbox{.}(2021)]%
        {bau2021paint}
\bibfield{author}{\bibinfo{person}{David Bau}, \bibinfo{person}{Alex Andonian}, \bibinfo{person}{Audrey Cui}, \bibinfo{person}{YeonHwan Park}, \bibinfo{person}{Ali Jahanian}, \bibinfo{person}{Aude Oliva}, {and} \bibinfo{person}{Antonio Torralba}.} \bibinfo{year}{2021}\natexlab{}.
\newblock \showarticletitle{Paint by word}.
\newblock \bibinfo{journal}{\emph{arXiv preprint arXiv:2103.10951}} (\bibinfo{year}{2021}).
\newblock


\bibitem[Brock et~al\mbox{.}(2019)]%
        {brock2019large}
\bibfield{author}{\bibinfo{person}{Andrew Brock}, \bibinfo{person}{Jeff Donahue}, {and} \bibinfo{person}{Karen Simonyan}.} \bibinfo{year}{2019}\natexlab{}.
\newblock \showarticletitle{Large Scale {GAN} Training for High Fidelity Natural Image Synthesis}. In \bibinfo{booktitle}{\emph{International Conference on Learning Representations (ICLR)}}.
\newblock


\bibitem[Brooks et~al\mbox{.}(2023)]%
        {brooks2022instructpix2pix}
\bibfield{author}{\bibinfo{person}{Tim Brooks}, \bibinfo{person}{Aleksander Holynski}, {and} \bibinfo{person}{Alexei~A Efros}.} \bibinfo{year}{2023}\natexlab{}.
\newblock \showarticletitle{{InstructPix2Pix}: Learning to Follow Image Editing Instructions}. In \bibinfo{booktitle}{\emph{IEEE/CVF Conference on Computer Vision and Pattern Recognition (CVPR)}}. \bibinfo{pages}{18392--18402}.
\newblock


\bibitem[Buss et~al\mbox{.}(1987a)]%
        {897367}
\bibfield{author}{\bibinfo{person}{Jonathan~F. Buss}, \bibinfo{person}{Arnold~L. Rosenberg}, {and} \bibinfo{person}{Judson~D. Knott}.} \bibinfo{year}{1987}\natexlab{a}.
\newblock \bibinfo{booktitle}{\emph{Vertex Types in Book-Embeddings}}.
\newblock \bibinfo{type}{{T}echnical {R}eport}. \bibinfo{address}{Amherst, MA, USA}.
\newblock


\bibitem[Buss et~al\mbox{.}(1987b)]%
        {Buss:1987:VTB:897367}
\bibfield{author}{\bibinfo{person}{Jonathan~F. Buss}, \bibinfo{person}{Arnold~L. Rosenberg}, {and} \bibinfo{person}{Judson~D. Knott}.} \bibinfo{year}{1987}\natexlab{b}.
\newblock \bibinfo{booktitle}{\emph{Vertex Types in Book-Embeddings}}.
\newblock \bibinfo{type}{{T}echnical {R}eport}. \bibinfo{address}{Amherst, MA, USA}.
\newblock


\bibitem[Chang et~al\mbox{.}(2023)]%
        {chang2023muse}
\bibfield{author}{\bibinfo{person}{Huiwen Chang}, \bibinfo{person}{Han Zhang}, \bibinfo{person}{Jarred Barber}, \bibinfo{person}{AJ Maschinot}, \bibinfo{person}{Jose Lezama}, \bibinfo{person}{Lu Jiang}, \bibinfo{person}{Ming-Hsuan Yang}, \bibinfo{person}{Kevin Murphy}, \bibinfo{person}{William~T. Freeman}, \bibinfo{person}{Michael Rubinstein}, \bibinfo{person}{Yuanzhen Li}, {and} \bibinfo{person}{Dilip Krishnan}.} \bibinfo{year}{2023}\natexlab{}.
\newblock \showarticletitle{Muse: Text-To-Image Generation via Masked Generative Transformers}. In \bibinfo{booktitle}{\emph{International Conference on Machine Learning (ICML)}}.
\newblock


\bibitem[Chong and Forsyth(2022)]%
        {chong2022jojogan}
\bibfield{author}{\bibinfo{person}{Min~Jin Chong} {and} \bibinfo{person}{David Forsyth}.} \bibinfo{year}{2022}\natexlab{}.
\newblock \showarticletitle{{JoJoGAN}: One Shot Face Stylization}. In \bibinfo{booktitle}{\emph{European Conference on Computer Vision (ECCV)}} (Tel Aviv, Israel). \bibinfo{publisher}{Springer-Verlag}, \bibinfo{address}{Berlin, Heidelberg}, \bibinfo{pages}{128–152}.
\newblock


\bibitem[Clarkson(1985a)]%
        {Clarkson85}
\bibfield{author}{\bibinfo{person}{Kenneth~L. Clarkson}.} \bibinfo{year}{1985}\natexlab{a}.
\newblock \emph{\bibinfo{title}{Algorithms for Closest-Point Problems (Computational Geometry)}}.
\newblock \bibinfo{thesistype}{Ph.\,D. Dissertation}. \bibinfo{school}{Stanford University}, \bibinfo{address}{Palo Alto, CA}.
\newblock
\newblock
\shownote{UMI Order Number: AAT 8506171}.


\bibitem[Clarkson(1985b)]%
        {Clarkson:1985:ACP:911891}
\bibfield{author}{\bibinfo{person}{Kenneth~Lee Clarkson}.} \bibinfo{year}{1985}\natexlab{b}.
\newblock \emph{\bibinfo{title}{Algorithms for Closest-Point Problems (Computational Geometry)}}.
\newblock \bibinfo{thesistype}{Ph.\,D. Dissertation}. \bibinfo{school}{Stanford University}, \bibinfo{address}{Stanford, CA, USA}.
\newblock Advisor(s) Yao, Andrew C.
\newblock
\newblock
\shownote{AAT 8506171}.


\bibitem[Cohen(1996)]%
        {JCohen96}
\bibfield{editor}{\bibinfo{person}{Jacques Cohen}} (Ed.). \bibinfo{year}{1996}\natexlab{}. \showarticletitle{Special issue: Digital Libraries}.
\newblock \bibinfo{journal}{\emph{Commun. {ACM}}} \bibinfo{volume}{39}, \bibinfo{number}{11} (\bibinfo{year}{1996}).

\bibitem[Cohen et~al\mbox{.}(2007)]%
        {Cohen07}
\bibfield{author}{\bibinfo{person}{Sarah Cohen}, \bibinfo{person}{Werner Nutt}, {and} \bibinfo{person}{Yehoshua Sagic}.} \bibinfo{year}{2007}\natexlab{}.
\newblock \showarticletitle{Deciding equivalances among conjunctive aggregate queries}.
\newblock \bibinfo{journal}{\emph{J. ACM}} \bibinfo{volume}{54}, \bibinfo{number}{2}, Article \bibinfo{articleno}{5} (\bibinfo{year}{2007}), \bibinfo{numpages}{50}~pages.
\newblock
\urldef\tempurl%
\url{https://doi.org/10.1145/1219092.1219093}
\showDOI{\tempurl}


\bibitem[Conti et~al\mbox{.}(2009a)]%
        {Conti:2009:DDS:1555009.1555162}
\bibfield{author}{\bibinfo{person}{Mauro Conti}, \bibinfo{person}{Roberto Di~Pietro}, \bibinfo{person}{Luigi~V. Mancini}, {and} \bibinfo{person}{Alessandro Mei}.} \bibinfo{year}{2009}\natexlab{a}.
\newblock \showarticletitle{(new) Distributed data source verification in wireless sensor networks}.
\newblock \bibinfo{journal}{\emph{Inf. Fusion}} \bibinfo{volume}{10}, \bibinfo{number}{4} (\bibinfo{year}{2009}), \bibinfo{pages}{342--353}.
\newblock
\showISSN{1566-2535}
\urldef\tempurl%
\url{https://doi.org/10.1016/j.inffus.2009.01.002}
\showDOI{\tempurl}


\bibitem[Conti et~al\mbox{.}(2009b)]%
        {1555162}
\bibfield{author}{\bibinfo{person}{Mauro Conti}, \bibinfo{person}{Roberto Di~Pietro}, \bibinfo{person}{Luigi~V. Mancini}, {and} \bibinfo{person}{Alessandro Mei}.} \bibinfo{year}{2009}\natexlab{b}.
\newblock \showarticletitle{(old) Distributed data source verification in wireless sensor networks}.
\newblock \bibinfo{journal}{\emph{Inf. Fusion}} \bibinfo{volume}{10}, \bibinfo{number}{4} (\bibinfo{year}{2009}), \bibinfo{pages}{342--353}.
\newblock
\showISSN{1566-2535}
\urldef\tempurl%
\url{https://doi.org/10.1016/j.inffus.2009.01.002}
\showDOI{\tempurl}


\bibitem[Crowson et~al\mbox{.}(2022)]%
        {crowson2022vqgan}
\bibfield{author}{\bibinfo{person}{Katherine Crowson}, \bibinfo{person}{Stella Biderman}, \bibinfo{person}{Daniel Kornis}, \bibinfo{person}{Dashiell Stander}, \bibinfo{person}{Eric Hallahan}, \bibinfo{person}{Louis Castricato}, {and} \bibinfo{person}{Edward Raff}.} \bibinfo{year}{2022}\natexlab{}.
\newblock \showarticletitle{{VQGAN-CLIP}: Open domain image generation and editing with natural language guidance}. In \bibinfo{booktitle}{\emph{European Conference on Computer Vision (ECCV)}}. Springer, \bibinfo{pages}{88--105}.
\newblock


\bibitem[Deng et~al\mbox{.}(2022)]%
        {deng2022stytr2}
\bibfield{author}{\bibinfo{person}{Yingying Deng}, \bibinfo{person}{Fan Tang}, \bibinfo{person}{Weiming Dong}, \bibinfo{person}{Chongyang Ma}, \bibinfo{person}{Xingjia Pan}, \bibinfo{person}{Lei Wang}, {and} \bibinfo{person}{Changsheng Xu}.} \bibinfo{year}{2022}\natexlab{}.
\newblock \showarticletitle{StyTr$^2$: Image Style Transfer with Transformers}. In \bibinfo{booktitle}{\emph{IEEE/CVF Conference on Computer Vision and Pattern Recognition (CVPR)}}. \bibinfo{pages}{11326--11336}.
\newblock


\bibitem[Dhariwal and Nichol(2021)]%
        {dhariwal2021diffusion}
\bibfield{author}{\bibinfo{person}{Prafulla Dhariwal} {and} \bibinfo{person}{Alexander Nichol}.} \bibinfo{year}{2021}\natexlab{}.
\newblock \showarticletitle{Diffusion models beat {GANs} on image synthesis}. In \bibinfo{booktitle}{\emph{Advances in Neural Information Processing Systems (NeurIPS}}. \bibinfo{pages}{8780--8794}.
\newblock


\bibitem[Douglass et~al\mbox{.}(1998)]%
        {Douglass98}
\bibfield{author}{\bibinfo{person}{Bruce~P. Douglass}, \bibinfo{person}{David Harel}, {and} \bibinfo{person}{Mark~B. Trakhtenbrot}.} \bibinfo{year}{1998}\natexlab{}.
\newblock \showarticletitle{Statecarts in use: structured analysis and object-orientation}.
\newblock In \bibinfo{booktitle}{\emph{Lectures on Embedded Systems}}, \bibfield{editor}{\bibinfo{person}{Grzegorz Rozenberg} {and} \bibinfo{person}{Frits~W. Vaandrager}} (Eds.). \bibinfo{series}{Lecture Notes in Computer Science}, Vol.~\bibinfo{volume}{1494}. \bibinfo{publisher}{Springer-Verlag}, \bibinfo{address}{London}, \bibinfo{pages}{368--394}.
\newblock
\urldef\tempurl%
\url{https://doi.org/10.1007/3-540-65193-4_29}
\showDOI{\tempurl}


\bibitem[Editor(2007)]%
        {Editor00}
\bibfield{editor}{\bibinfo{person}{Ian Editor}} (Ed.). \bibinfo{year}{2007}\natexlab{}.
\newblock \bibinfo{booktitle}{\emph{The title of book one} (\bibinfo{edition}{1st.} ed.)}. \bibinfo{series}{The name of the series one}, Vol.~\bibinfo{volume}{9}.
\newblock \bibinfo{publisher}{University of Chicago Press}, \bibinfo{address}{Chicago}.
\newblock
\urldef\tempurl%
\url{https://doi.org/10.1007/3-540-09237-4}
\showDOI{\tempurl}


\bibitem[Editor(2008)]%
        {Editor00a}
\bibfield{editor}{\bibinfo{person}{Ian Editor}} (Ed.). \bibinfo{year}{2008}\natexlab{}.
\newblock \bibinfo{booktitle}{\emph{The title of book two} (\bibinfo{edition}{2nd.} ed.)}.
\newblock \bibinfo{publisher}{University of Chicago Press}, \bibinfo{address}{Chicago}, Chapter 100.
\newblock
\urldef\tempurl%
\url{https://doi.org/10.1007/3-540-09237-4}
\showDOI{\tempurl}


\bibitem[Esser et~al\mbox{.}(2021)]%
        {esser2021taming}
\bibfield{author}{\bibinfo{person}{Patrick Esser}, \bibinfo{person}{Robin Rombach}, {and} \bibinfo{person}{Bjorn Ommer}.} \bibinfo{year}{2021}\natexlab{}.
\newblock \showarticletitle{Taming Transformers for High-Resolution Image Synthesis}. In \bibinfo{booktitle}{\emph{IEEE/CVF Conference on Computer Vision and Pattern Recognition (CVPR)}}. \bibinfo{pages}{12873--12883}.
\newblock


\bibitem[Gafni et~al\mbox{.}(2022)]%
        {gafni2022make}
\bibfield{author}{\bibinfo{person}{Oran Gafni}, \bibinfo{person}{Adam Polyak}, \bibinfo{person}{Oron Ashual}, \bibinfo{person}{Shelly Sheynin}, \bibinfo{person}{Devi Parikh}, {and} \bibinfo{person}{Yaniv Taigman}.} \bibinfo{year}{2022}\natexlab{}.
\newblock \showarticletitle{Make-a-scene: Scene-based text-to-image generation with human priors}. In \bibinfo{booktitle}{\emph{European Conference on Computer Vision (ECCV)}}. Springer, \bibinfo{pages}{89--106}.
\newblock


\bibitem[Gal et~al\mbox{.}(2023a)]%
        {gal2022TI}
\bibfield{author}{\bibinfo{person}{Rinon Gal}, \bibinfo{person}{Yuval Alaluf}, \bibinfo{person}{Yuval Atzmon}, \bibinfo{person}{Or Patashnik}, \bibinfo{person}{Amit~H Bermano}, \bibinfo{person}{Gal Chechik}, {and} \bibinfo{person}{Daniel Cohen-Or}.} \bibinfo{year}{2023}\natexlab{a}.
\newblock \showarticletitle{An Image is Worth One Word: Personalizing Text-to-Image Generation using Textual Inversion}. In \bibinfo{booktitle}{\emph{International Conference on Learning Representations (ICLR)}}.
\newblock


\bibitem[Gal et~al\mbox{.}(2023b)]%
        {gal2023encoder}
\bibfield{author}{\bibinfo{person}{Rinon Gal}, \bibinfo{person}{Moab Arar}, \bibinfo{person}{Yuval Atzmon}, \bibinfo{person}{Amit~H Bermano}, \bibinfo{person}{Gal Chechik}, {and} \bibinfo{person}{Daniel Cohen-Or}.} \bibinfo{year}{2023}\natexlab{b}.
\newblock \showarticletitle{Encoder-based domain tuning for fast personalization of text-to-image models}.
\newblock \bibinfo{journal}{\emph{ACM Transactions on Graphics (TOG)}} \bibinfo{volume}{42}, \bibinfo{number}{4} (\bibinfo{year}{2023}), \bibinfo{pages}{1--13}.
\newblock


\bibitem[Gal et~al\mbox{.}(2022)]%
        {StyleGANNADA}
\bibfield{author}{\bibinfo{person}{Rinon Gal}, \bibinfo{person}{Or Patashnik}, \bibinfo{person}{Haggai Maron}, \bibinfo{person}{Amit~H. Bermano}, \bibinfo{person}{Gal Chechik}, {and} \bibinfo{person}{Daniel Cohen-Or}.} \bibinfo{year}{2022}\natexlab{}.
\newblock \showarticletitle{{StyleGAN-NADA: CLIP}-Guided Domain Adaptation of Image Generators}.
\newblock \bibinfo{journal}{\emph{ACM Transactions on Graphics}} \bibinfo{volume}{41}, \bibinfo{number}{4}, Article \bibinfo{articleno}{141} (\bibinfo{year}{2022}), \bibinfo{numpages}{13}~pages.
\newblock


\bibitem[Geiger and Meek(2005)]%
        {GM05}
\bibfield{author}{\bibinfo{person}{Dan Geiger} {and} \bibinfo{person}{Christopher Meek}.} \bibinfo{year}{2005}\natexlab{}.
\newblock \showarticletitle{Structured Variational Inference Procedures and their Realizations (as incol)}.
\newblock In \bibinfo{booktitle}{\emph{Proceedings of Tenth International Workshop on Artificial Intelligence and Statistics, {\rm The Barbados}}}. \bibinfo{publisher}{The Society for Artificial Intelligence and Statistics}.
\newblock


\bibitem[Goodfellow et~al\mbox{.}(2014)]%
        {Goodfellow:2014:GAN}
\bibfield{author}{\bibinfo{person}{Ian Goodfellow}, \bibinfo{person}{Jean Pouget-Abadie}, \bibinfo{person}{Mehdi Mirza}, \bibinfo{person}{Bing Xu}, \bibinfo{person}{David Warde-Farley}, \bibinfo{person}{Sherjil Ozair}, \bibinfo{person}{Aaron Courville}, {and} \bibinfo{person}{Yoshua Bengio}.} \bibinfo{year}{2014}\natexlab{}.
\newblock \showarticletitle{Generative Adversarial Nets}. In \bibinfo{booktitle}{\emph{Advances in Neural Information Processing Systems (NIPS)}}. \bibinfo{publisher}{Curran Associates, Inc.}
\newblock


\bibitem[Goossens et~al\mbox{.}(1999)]%
        {Goossens:1999:LWC:553897}
\bibfield{author}{\bibinfo{person}{Michel Goossens}, \bibinfo{person}{S.~P. Rahtz}, \bibinfo{person}{Ross Moore}, {and} \bibinfo{person}{Robert~S. Sutor}.} \bibinfo{year}{1999}\natexlab{}.
\newblock \bibinfo{booktitle}{\emph{The Latex Web Companion: Integrating TEX, HTML, and XML} (\bibinfo{edition}{1st} ed.)}.
\newblock \bibinfo{publisher}{Addison-Wesley Longman Publishing Co., Inc.}, \bibinfo{address}{Boston, MA, USA}.
\newblock
\showISBNx{0201433117}


\bibitem[Gundy et~al\mbox{.}(2007)]%
        {VanGundy07}
\bibfield{author}{\bibinfo{person}{Matthew~Van Gundy}, \bibinfo{person}{Davide Balzarotti}, {and} \bibinfo{person}{Giovanni Vigna}.} \bibinfo{year}{2007}\natexlab{}.
\newblock \showarticletitle{Catch me, if you can: Evading network signatures with web-based polymorphic worms}. In \bibinfo{booktitle}{\emph{Proceedings of the first USENIX workshop on Offensive Technologies}} \emph{(\bibinfo{series}{WOOT '07})}. \bibinfo{publisher}{USENIX Association}, \bibinfo{address}{Berkley, CA}, Article \bibinfo{articleno}{7}, \bibinfo{numpages}{9}~pages.
\newblock


\bibitem[Gundy et~al\mbox{.}(2008)]%
        {VanGundy08}
\bibfield{author}{\bibinfo{person}{Matthew~Van Gundy}, \bibinfo{person}{Davide Balzarotti}, {and} \bibinfo{person}{Giovanni Vigna}.} \bibinfo{year}{2008}\natexlab{}.
\newblock \showarticletitle{Catch me, if you can: Evading network signatures with web-based polymorphic worms}. In \bibinfo{booktitle}{\emph{Proceedings of the first USENIX workshop on Offensive Technologies}} \emph{(\bibinfo{series}{WOOT '08})}. \bibinfo{publisher}{USENIX Association}, \bibinfo{address}{Berkley, CA}, Article \bibinfo{articleno}{7}, \bibinfo{numpages}{2}~pages.
\newblock


\bibitem[Gundy et~al\mbox{.}(2009)]%
        {VanGundy09}
\bibfield{author}{\bibinfo{person}{Matthew~Van Gundy}, \bibinfo{person}{Davide Balzarotti}, {and} \bibinfo{person}{Giovanni Vigna}.} \bibinfo{year}{2009}\natexlab{}.
\newblock \showarticletitle{Catch me, if you can: Evading network signatures with web-based polymorphic worms}. In \bibinfo{booktitle}{\emph{Proceedings of the first USENIX workshop on Offensive Technologies}} \emph{(\bibinfo{series}{WOOT '09})}. \bibinfo{publisher}{USENIX Association}, \bibinfo{address}{Berkley, CA}, \bibinfo{pages}{90--100}.
\newblock


\bibitem[Harel(1978)]%
        {Harel78}
\bibfield{author}{\bibinfo{person}{David Harel}.} \bibinfo{year}{1978}\natexlab{}.
\newblock \bibinfo{booktitle}{\emph{LOGICS of Programs: AXIOMATICS and DESCRIPTIVE POWER}}.
\newblock \bibinfo{type}{MIT Research Lab Technical Report} TR-200. \bibinfo{institution}{Massachusetts Institute of Technology}, \bibinfo{address}{Cambridge, MA}.
\newblock


\bibitem[Harel(1979)]%
        {Harel79}
\bibfield{author}{\bibinfo{person}{David Harel}.} \bibinfo{year}{1979}\natexlab{}.
\newblock \bibinfo{booktitle}{\emph{First-Order Dynamic Logic}}. \bibinfo{series}{Lecture Notes in Computer Science}, Vol.~\bibinfo{volume}{68}.
\newblock \bibinfo{publisher}{Springer-Verlag}, \bibinfo{address}{New York, NY}.
\newblock
\urldef\tempurl%
\url{https://doi.org/10.1007/3-540-09237-4}
\showDOI{\tempurl}


\bibitem[Hertz et~al\mbox{.}(2023)]%
        {hertz2022p2p}
\bibfield{author}{\bibinfo{person}{Amir Hertz}, \bibinfo{person}{Ron Mokady}, \bibinfo{person}{Jay Tenenbaum}, \bibinfo{person}{Kfir Aberman}, \bibinfo{person}{Yael Pritch}, {and} \bibinfo{person}{Daniel Cohen-Or}.} \bibinfo{year}{2023}\natexlab{}.
\newblock \showarticletitle{Prompt-to-Prompt Image Editing with Cross Attention Control}. In \bibinfo{booktitle}{\emph{International Conference on Learning Representations (ICLR)}}.
\newblock


\bibitem[Hertzmann et~al\mbox{.}(2001)]%
        {hertzmann2001analogies}
\bibfield{author}{\bibinfo{person}{Aaron Hertzmann}, \bibinfo{person}{Charles~E Jacobs}, \bibinfo{person}{Nuria Oliver}, \bibinfo{person}{Brian Curless}, {and} \bibinfo{person}{David~H Salesin}.} \bibinfo{year}{2001}\natexlab{}.
\newblock \showarticletitle{Image analogies}. In \bibinfo{booktitle}{\emph{Proceedings of the 28th annual conference on Computer graphics and interactive techniques}}. \bibinfo{pages}{327--340}.
\newblock


\bibitem[Heusel et~al\mbox{.}(2017)]%
        {heusel2017gans}
\bibfield{author}{\bibinfo{person}{Martin Heusel}, \bibinfo{person}{Hubert Ramsauer}, \bibinfo{person}{Thomas Unterthiner}, \bibinfo{person}{Bernhard Nessler}, {and} \bibinfo{person}{Sepp Hochreiter}.} \bibinfo{year}{2017}\natexlab{}.
\newblock \showarticletitle{{GANs} Trained by a Two Time-Scale Update Rule Converge to a Local Nash Equilibrium}. In \bibinfo{booktitle}{\emph{Advances in Neural Information Processing Systems (NIPS)}}.
\newblock


\bibitem[Hollis(1999)]%
        {Hollis:1999:VBD:519964}
\bibfield{author}{\bibinfo{person}{Billy~S. Hollis}.} \bibinfo{year}{1999}\natexlab{}.
\newblock \bibinfo{booktitle}{\emph{Visual Basic 6: Design, Specification, and Objects with Other} (\bibinfo{edition}{1st} ed.)}.
\newblock \bibinfo{publisher}{Prentice Hall PTR}, \bibinfo{address}{Upper Saddle River, NJ, USA}.
\newblock
\showISBNx{0130850845}


\bibitem[Huang et~al\mbox{.}(2023a)]%
        {huang2023composer}
\bibfield{author}{\bibinfo{person}{Lianghua Huang}, \bibinfo{person}{Di Chen}, \bibinfo{person}{Yu Liu}, \bibinfo{person}{Yujun Shen}, \bibinfo{person}{Deli Zhao}, {and} \bibinfo{person}{Jingren Zhou}.} \bibinfo{year}{2023}\natexlab{a}.
\newblock \showarticletitle{Composer: Creative and Controllable Image Synthesis with Composable Conditions}. In \bibinfo{booktitle}{\emph{International Conference on Machine Learning (ICML)}}.
\newblock


\bibitem[Huang et~al\mbox{.}(2023b)]%
        {lhhuang2023composer}
\bibfield{author}{\bibinfo{person}{Lianghua Huang}, \bibinfo{person}{Di Chen}, \bibinfo{person}{Yu Liu}, \bibinfo{person}{Shen Yujun}, \bibinfo{person}{Deli Zhao}, {and} \bibinfo{person}{Zhou Jingren}.} \bibinfo{year}{2023}\natexlab{b}.
\newblock \showarticletitle{Composer: Creative and Controllable Image Synthesis with Composable Conditions}.
\newblock  (\bibinfo{year}{2023}).
\newblock


\bibitem[Huang et~al\mbox{.}(2023c)]%
        {huang2023region}
\bibfield{author}{\bibinfo{person}{Nisha Huang}, \bibinfo{person}{Fan Tang}, \bibinfo{person}{Weiming Dong}, \bibinfo{person}{Tong-Yee Lee}, {and} \bibinfo{person}{Changsheng Xu}.} \bibinfo{year}{2023}\natexlab{c}.
\newblock \showarticletitle{Region-Aware Diffusion for Zero-shot Text-driven Image Editing}.
\newblock \bibinfo{journal}{\emph{arXiv preprint arXiv:2302.11797}} (\bibinfo{year}{2023}).
\newblock


\bibitem[Huang et~al\mbox{.}(2022a)]%
        {Huang2022MGAD}
\bibfield{author}{\bibinfo{person}{Nisha Huang}, \bibinfo{person}{Fan Tang}, \bibinfo{person}{Weiming Dong}, {and} \bibinfo{person}{Changsheng Xu}.} \bibinfo{year}{2022}\natexlab{a}.
\newblock \showarticletitle{Draw Your Art Dream: Diverse Digital Art Synthesis with Multimodal Guided Diffusion}. In \bibinfo{booktitle}{\emph{ACM International Conference on Multimedia}} (Lisboa, Portugal). \bibinfo{pages}{1085–1094}.
\newblock


\bibitem[Huang et~al\mbox{.}(2023e)]%
        {huang2023style}
\bibfield{author}{\bibinfo{person}{Nisha Huang}, \bibinfo{person}{Yuxin Zhang}, {and} \bibinfo{person}{Weiming Dong}.} \bibinfo{year}{2023}\natexlab{e}.
\newblock \showarticletitle{Style-A-Video: Agile Diffusion for Arbitrary Text-based Video Style Transfer}.
\newblock \bibinfo{journal}{\emph{arXiv preprint arXiv:2305.05464}} (\bibinfo{year}{2023}).
\newblock


\bibitem[Huang et~al\mbox{.}(2022b)]%
        {huang2022diffstyler}
\bibfield{author}{\bibinfo{person}{Nisha Huang}, \bibinfo{person}{Yuxin Zhang}, \bibinfo{person}{Fan Tang}, \bibinfo{person}{Chongyang Ma}, \bibinfo{person}{Haibin Huang}, \bibinfo{person}{Yong Zhang}, \bibinfo{person}{Weiming Dong}, {and} \bibinfo{person}{Changsheng Xu}.} \bibinfo{year}{2022}\natexlab{b}.
\newblock \showarticletitle{DiffStyler: Controllable Dual Diffusion for Text-Driven Image Stylization}.
\newblock \bibinfo{journal}{\emph{arXiv preprint arXiv:2211.10682}} (\bibinfo{year}{2022}).
\newblock


\bibitem[Huang et~al\mbox{.}(2018)]%
        {munit}
\bibfield{author}{\bibinfo{person}{Xun Huang}, \bibinfo{person}{Ming-Yu Liu}, \bibinfo{person}{Serge Belongie}, {and} \bibinfo{person}{Jan Kautz}.} \bibinfo{year}{2018}\natexlab{}.
\newblock \showarticletitle{Multimodal Unsupervised Image-to-Image Translation}. In \bibinfo{booktitle}{\emph{European Conference on Computer Vision (ECCV)}}. \bibinfo{pages}{172--189}.
\newblock


\bibitem[Huang et~al\mbox{.}(2023d)]%
        {huang2023reversion}
\bibfield{author}{\bibinfo{person}{Ziqi Huang}, \bibinfo{person}{Tianxing Wu}, \bibinfo{person}{Yuming Jiang}, \bibinfo{person}{Kelvin~CK Chan}, {and} \bibinfo{person}{Ziwei Liu}.} \bibinfo{year}{2023}\natexlab{d}.
\newblock \showarticletitle{ReVersion: Diffusion-Based Relation Inversion from Images}.
\newblock \bibinfo{journal}{\emph{arXiv preprint arXiv:2303.13495}} (\bibinfo{year}{2023}).
\newblock


\bibitem[IEEE(2004)]%
        {2004:ITE:1009386.1010128}
IEEE \bibinfo{year}{2004}\natexlab{}.
\newblock \showarticletitle{IEEE TCSC Executive Committee}. In \bibinfo{booktitle}{\emph{Proceedings of the IEEE International Conference on Web Services}} \emph{(\bibinfo{series}{ICWS '04})}. \bibinfo{publisher}{IEEE Computer Society}, \bibinfo{address}{Washington, DC, USA}, \bibinfo{pages}{21--22}.
\newblock
\showISBNx{0-7695-2167-3}
\urldef\tempurl%
\url{https://doi.org/10.1109/ICWS.2004.64}
\showDOI{\tempurl}


\bibitem[Jeong et~al\mbox{.}(2023)]%
        {jeong2023training}
\bibfield{author}{\bibinfo{person}{Jaeseok Jeong}, \bibinfo{person}{Mingi Kwon}, {and} \bibinfo{person}{Youngjung Uh}.} \bibinfo{year}{2023}\natexlab{}.
\newblock \showarticletitle{Training-free Style Transfer Emerges from h-space in Diffusion models}.
\newblock \bibinfo{journal}{\emph{arXiv preprint arXiv:2303.15403}} (\bibinfo{year}{2023}).
\newblock


\bibitem[Karras et~al\mbox{.}(2020)]%
        {karras2020training}
\bibfield{author}{\bibinfo{person}{Tero Karras}, \bibinfo{person}{Miika Aittala}, \bibinfo{person}{Janne Hellsten}, \bibinfo{person}{Samuli Laine}, \bibinfo{person}{Jaakko Lehtinen}, {and} \bibinfo{person}{Timo Aila}.} \bibinfo{year}{2020}\natexlab{}.
\newblock \showarticletitle{Training Generative Adversarial Networks with Limited Data}. In \bibinfo{booktitle}{\emph{Advances in Neural Information Processing Systems (NeurIPS)}}. \bibinfo{pages}{12104--12114}.
\newblock


\bibitem[Karras et~al\mbox{.}(2019)]%
        {stylegan}
\bibfield{author}{\bibinfo{person}{Tero Karras}, \bibinfo{person}{Samuli Laine}, {and} \bibinfo{person}{Timo Aila}.} \bibinfo{year}{2019}\natexlab{}.
\newblock \showarticletitle{A Style-Based Generator Architecture for Generative Adversarial Networks}. In \bibinfo{booktitle}{\emph{IEEE/CVF Conference on Computer Vision and Pattern Recognition (CVPR)}}. \bibinfo{pages}{4401--4410}.
\newblock


\bibitem[Kawar et~al\mbox{.}(2023)]%
        {imagic}
\bibfield{author}{\bibinfo{person}{Bahjat Kawar}, \bibinfo{person}{Shiran Zada}, \bibinfo{person}{Oran Lang}, \bibinfo{person}{Omer Tov}, \bibinfo{person}{Huiwen Chang}, \bibinfo{person}{Tali Dekel}, \bibinfo{person}{Inbar Mosseri}, {and} \bibinfo{person}{Michal Irani}.} \bibinfo{year}{2023}\natexlab{}.
\newblock \showarticletitle{Imagic: Text-Based Real Image Editing with Diffusion Models}. In \bibinfo{booktitle}{\emph{IEEE/CVF Conference on Computer Vision and Pattern Recognition (CVPR)}}. \bibinfo{pages}{6007--6017}.
\newblock


\bibitem[Knuth(1981)]%
        {book-minimal}
\bibfield{author}{\bibinfo{person}{Donald~E. Knuth}.} \bibinfo{year}{1981}\natexlab{}.
\newblock \bibinfo{booktitle}{\emph{Seminumerical Algorithms}}.
\newblock \bibinfo{publisher}{Addison-Wesley}.
\newblock


\bibitem[Knuth(1997)]%
        {Knuth97}
\bibfield{author}{\bibinfo{person}{Donald~E. Knuth}.} \bibinfo{year}{1997}\natexlab{}.
\newblock \bibinfo{booktitle}{\emph{The Art of Computer Programming, Vol. 1: Fundamental Algorithms (3rd. ed.)}}.
\newblock \bibinfo{publisher}{Addison Wesley Longman Publishing Co., Inc.}
\newblock


\bibitem[Knuth(1998)]%
        {Knuth98}
\bibfield{author}{\bibinfo{person}{Donald~E. Knuth}.} \bibinfo{year}{1998}\natexlab{}.
\newblock \bibinfo{booktitle}{\emph{The Art of Computer Programming} (\bibinfo{edition}{3rd} ed.)}. \bibinfo{series}{Fundamental Algorithms}, Vol.~\bibinfo{volume}{1}.
\newblock \bibinfo{publisher}{Addison Wesley Longman Publishing Co., Inc.}
\newblock
\newblock
\shownote{(book)}.


\bibitem[Kong(2001a)]%
        {KAGM:2001}
\bibfield{author}{\bibinfo{person}{Wei-Chang Kong}.} \bibinfo{year}{2001}\natexlab{a}.
\newblock \bibinfo{booktitle}{\emph{E-commerce and cultural values}}.
\newblock \bibinfo{publisher}{IGI Publishing}, \bibinfo{address}{Hershey, PA, USA}, Name of chapter: The implementation of electronic commerce in SMEs in Singapore (Inbook-w-chap-w-type), \bibinfo{pages}{51--74}.
\newblock
\showISBNx{1-59140-056-2}
\urldef\tempurl%
\url{http://portal.acm.org/citation.cfm?id=887006.887010}
\showURL{%
\tempurl}


\bibitem[Kong(2001b)]%
        {KA:2001}
\bibfield{author}{\bibinfo{person}{Wei-Chang Kong}.} \bibinfo{year}{2001}\natexlab{b}.
\newblock \showarticletitle{The implementation of electronic commerce in SMEs in Singapore (as Incoll)}.
\newblock In \bibinfo{booktitle}{\emph{E-commerce and cultural values}}. \bibinfo{publisher}{IGI Publishing}, \bibinfo{address}{Hershey, PA, USA}, \bibinfo{pages}{51--74}.
\newblock
\showISBNx{1-59140-056-2}
\urldef\tempurl%
\url{http://portal.acm.org/citation.cfm?id=887006.887010}
\showURL{%
\tempurl}


\bibitem[Kong(2002)]%
        {Kong:2002:IEC:887006.887010}
\bibfield{author}{\bibinfo{person}{Wei-Chang Kong}.} \bibinfo{year}{2002}\natexlab{}.
\newblock \showarticletitle{Chapter 9}.
\newblock In \bibinfo{booktitle}{\emph{E-commerce and cultural values (Incoll-w-text (chap 9) 'title')}}, \bibfield{editor}{\bibinfo{person}{Theerasak Thanasankit}} (Ed.). \bibinfo{publisher}{IGI Publishing}, \bibinfo{address}{Hershey, PA, USA}, \bibinfo{pages}{51--74}.
\newblock
\showISBNx{1-59140-056-2}
\urldef\tempurl%
\url{http://portal.acm.org/citation.cfm?id=887006.887010}
\showURL{%
\tempurl}


\bibitem[Kong(2003)]%
        {Kong:2003:IEC:887006.887011}
\bibfield{author}{\bibinfo{person}{Wei-Chang Kong}.} \bibinfo{year}{2003}\natexlab{}.
\newblock \showarticletitle{The implementation of electronic commerce in SMEs in Singapore (Incoll)}.
\newblock In \bibinfo{booktitle}{\emph{E-commerce and cultural values}}, \bibfield{editor}{\bibinfo{person}{Theerasak Thanasankit}} (Ed.). \bibinfo{publisher}{IGI Publishing}, \bibinfo{address}{Hershey, PA, USA}, \bibinfo{pages}{51--74}.
\newblock
\showISBNx{1-59140-056-2}
\urldef\tempurl%
\url{http://portal.acm.org/citation.cfm?id=887006.887010}
\showURL{%
\tempurl}


\bibitem[Kong(2004)]%
        {Kong:2004:IEC:123456.887010}
\bibfield{author}{\bibinfo{person}{Wei-Chang Kong}.} \bibinfo{year}{2004}\natexlab{}.
\newblock \bibinfo{booktitle}{\emph{E-commerce and cultural values - (InBook-num-in-chap)}}.
\newblock \bibinfo{publisher}{IGI Publishing}, \bibinfo{address}{Hershey, PA, USA}, Chapter~9, \bibinfo{pages}{51--74}.
\newblock
\showISBNx{1-59140-056-2}
\urldef\tempurl%
\url{http://portal.acm.org/citation.cfm?id=887006.887010}
\showURL{%
\tempurl}


\bibitem[Kong(2005)]%
        {Kong:2005:IEC:887006.887010}
\bibfield{author}{\bibinfo{person}{Wei-Chang Kong}.} \bibinfo{year}{2005}\natexlab{}.
\newblock \bibinfo{booktitle}{\emph{E-commerce and cultural values (Inbook-text-in-chap)}}.
\newblock \bibinfo{publisher}{IGI Publishing}, \bibinfo{address}{Hershey, PA, USA}, Chapter: The implementation of electronic commerce in SMEs in Singapore, \bibinfo{pages}{51--74}.
\newblock
\showISBNx{1-59140-056-2}
\urldef\tempurl%
\url{http://portal.acm.org/citation.cfm?id=887006.887010}
\showURL{%
\tempurl}


\bibitem[Kong(2006)]%
        {Kong:2006:IEC:887006.887010}
\bibfield{author}{\bibinfo{person}{Wei-Chang Kong}.} \bibinfo{year}{2006}\natexlab{}.
\newblock \bibinfo{booktitle}{\emph{E-commerce and cultural values (Inbook-num chap)}}.
\newblock \bibinfo{publisher}{IGI Publishing}, \bibinfo{address}{Hershey, PA, USA}, Chapter (in type field)~22, \bibinfo{pages}{51--74}.
\newblock
\showISBNx{1-59140-056-2}
\urldef\tempurl%
\url{http://portal.acm.org/citation.cfm?id=887006.887010}
\showURL{%
\tempurl}


\bibitem[Kosiur(2001)]%
        {Kosiur01}
\bibfield{author}{\bibinfo{person}{David Kosiur}.} \bibinfo{year}{2001}\natexlab{}.
\newblock \bibinfo{booktitle}{\emph{Understanding Policy-Based Networking} (\bibinfo{edition}{2nd.} ed.)}.
\newblock \bibinfo{publisher}{Wiley}, \bibinfo{address}{New York, NY}.
\newblock


\bibitem[Kumari et~al\mbox{.}(2023a)]%
        {kumari2022multi}
\bibfield{author}{\bibinfo{person}{Nupur Kumari}, \bibinfo{person}{Bingliang Zhang}, \bibinfo{person}{Richard Zhang}, \bibinfo{person}{Eli Shechtman}, {and} \bibinfo{person}{Jun-Yan Zhu}.} \bibinfo{year}{2023}\natexlab{a}.
\newblock \showarticletitle{Multi-Concept Customization of Text-to-Image Diffusion}. In \bibinfo{booktitle}{\emph{IEEE/CVF Conference on Computer Vision and Pattern Recognition (CVPR)}}. \bibinfo{pages}{1931--1941}.
\newblock


\bibitem[Kumari et~al\mbox{.}(2023b)]%
        {kumari2022customdiffusion}
\bibfield{author}{\bibinfo{person}{Nupur Kumari}, \bibinfo{person}{Bingliang Zhang}, \bibinfo{person}{Richard Zhang}, \bibinfo{person}{Eli Shechtman}, {and} \bibinfo{person}{Jun-Yan Zhu}.} \bibinfo{year}{2023}\natexlab{b}.
\newblock \showarticletitle{Multi-Concept Customization of Text-to-Image Diffusion}. In \bibinfo{booktitle}{\emph{IEEE/CVF Conference on Computer Vision and Pattern Recognition (CVPR)}}.
\newblock


\bibitem[Kwon and Ye(2022)]%
        {CLIPstyler}
\bibfield{author}{\bibinfo{person}{Gihyun Kwon} {and} \bibinfo{person}{Jong~Chul Ye}.} \bibinfo{year}{2022}\natexlab{}.
\newblock \showarticletitle{{CLIPstyler}: Image Style Transfer with a Single Text Condition}. In \bibinfo{booktitle}{\emph{IEEE/CVF Conference on Computer Vision and Pattern Recognition (CVPR)}}. \bibinfo{pages}{18062–18071}.
\newblock


\bibitem[Lee et~al\mbox{.}(2020)]%
        {lee2020drit}
\bibfield{author}{\bibinfo{person}{Hsin-Ying Lee}, \bibinfo{person}{Hung-Yu Tseng}, \bibinfo{person}{Qi Mao}, \bibinfo{person}{Jia-Bin Huang}, \bibinfo{person}{Yu-Ding Lu}, \bibinfo{person}{Maneesh Singh}, {and} \bibinfo{person}{Ming-Hsuan Yang}.} \bibinfo{year}{2020}\natexlab{}.
\newblock \showarticletitle{Drit++: Diverse image-to-image translation via disentangled representations}.
\newblock \bibinfo{journal}{\emph{International Journal of Computer Vision}}  \bibinfo{volume}{128} (\bibinfo{year}{2020}), \bibinfo{pages}{2402--2417}.
\newblock


\bibitem[Lee(2005)]%
        {Lee05}
\bibfield{author}{\bibinfo{person}{Newton Lee}.} \bibinfo{year}{2005}\natexlab{}.
\newblock \showarticletitle{Interview with Bill Kinder: January 13, 2005}.
\newblock \bibinfo{howpublished}{Video}.
\newblock \bibinfo{journal}{\emph{Comput. Entertain.}} \bibinfo{volume}{3}, \bibinfo{number}{1}, Article \bibinfo{articleno}{4} (\bibinfo{year}{2005}).
\newblock
\urldef\tempurl%
\url{https://doi.org/10.1145/1057270.1057278}
\showDOI{\tempurl}


\bibitem[Li et~al\mbox{.}(2008)]%
        {Li:2008:PUC:1358628.1358946}
\bibfield{author}{\bibinfo{person}{Cheng-Lun Li}, \bibinfo{person}{Ayse~G. Buyuktur}, \bibinfo{person}{David~K. Hutchful}, \bibinfo{person}{Natasha~B. Sant}, {and} \bibinfo{person}{Satyendra~K. Nainwal}.} \bibinfo{year}{2008}\natexlab{}.
\newblock \showarticletitle{Portalis: using competitive online interactions to support aid initiatives for the homeless}. In \bibinfo{booktitle}{\emph{CHI '08 extended abstracts on Human factors in computing systems}} (Florence, Italy). \bibinfo{publisher}{ACM}, \bibinfo{address}{New York, NY, USA}, \bibinfo{pages}{3873--3878}.
\newblock
\showISBNx{978-1-60558-012-X}
\urldef\tempurl%
\url{https://doi.org/10.1145/1358628.1358946}
\showDOI{\tempurl}


\bibitem[Li et~al\mbox{.}(2023)]%
        {li2023stylediffusion}
\bibfield{author}{\bibinfo{person}{Senmao Li}, \bibinfo{person}{Joost van~de Weijer}, \bibinfo{person}{Taihang Hu}, \bibinfo{person}{Fahad~Shahbaz Khan}, \bibinfo{person}{Qibin Hou}, \bibinfo{person}{Yaxing Wang}, {and} \bibinfo{person}{Jian Yang}.} \bibinfo{year}{2023}\natexlab{}.
\newblock \showarticletitle{{StyleDiffusion:} Prompt-Embedding Inversion for Text-Based Editing}.
\newblock \bibinfo{journal}{\emph{arXiv preprint arXiv:2303.15649}} (\bibinfo{year}{2023}).
\newblock


\bibitem[Liao et~al\mbox{.}(2022)]%
        {liao2022text}
\bibfield{author}{\bibinfo{person}{Wentong Liao}, \bibinfo{person}{Kai Hu}, \bibinfo{person}{Michael~Ying Yang}, {and} \bibinfo{person}{Bodo Rosenhahn}.} \bibinfo{year}{2022}\natexlab{}.
\newblock \showarticletitle{Text to Image Generation with Semantic-Spatial Aware GAN}. In \bibinfo{booktitle}{\emph{IEEE/CVF Conference on Computer Vision and Pattern Recognition (CVPR)}}. \bibinfo{pages}{18187--18196}.
\newblock


\bibitem[Lugmayr et~al\mbox{.}(2022)]%
        {lugmayr2022repaint}
\bibfield{author}{\bibinfo{person}{Andreas Lugmayr}, \bibinfo{person}{Martin Danelljan}, \bibinfo{person}{Andres Romero}, \bibinfo{person}{Fisher Yu}, \bibinfo{person}{Radu Timofte}, {and} \bibinfo{person}{Luc Van~Gool}.} \bibinfo{year}{2022}\natexlab{}.
\newblock \showarticletitle{{RePaint:} Inpainting Using Denoising Diffusion Probabilistic Models}. In \bibinfo{booktitle}{\emph{IEEE/CVF Conference on Computer Vision and Pattern Recognition (CVPR)}}. \bibinfo{pages}{11461--11471}.
\newblock


\bibitem[Meng et~al\mbox{.}(2021)]%
        {meng2021sdedit}
\bibfield{author}{\bibinfo{person}{Chenlin Meng}, \bibinfo{person}{Yutong He}, \bibinfo{person}{Yang Song}, \bibinfo{person}{Jiaming Song}, \bibinfo{person}{Jiajun Wu}, \bibinfo{person}{Jun-Yan Zhu}, {and} \bibinfo{person}{Stefano Ermon}.} \bibinfo{year}{2021}\natexlab{}.
\newblock \showarticletitle{Sdedit: Guided image synthesis and editing with stochastic differential equations}.
\newblock \bibinfo{journal}{\emph{arXiv preprint arXiv:2108.01073}} (\bibinfo{year}{2021}).
\newblock


\bibitem[Meng et~al\mbox{.}(2022)]%
        {meng2022locating}
\bibfield{author}{\bibinfo{person}{Kevin Meng}, \bibinfo{person}{David Bau}, \bibinfo{person}{Alex Andonian}, {and} \bibinfo{person}{Yonatan Belinkov}.} \bibinfo{year}{2022}\natexlab{}.
\newblock \showarticletitle{Locating and editing factual associations in GPT}. In \bibinfo{booktitle}{\emph{Advances in Neural Information Processing Systems (NeurIPS)}}. \bibinfo{pages}{17359--17372}.
\newblock


\bibitem[Mokady et~al\mbox{.}(2023)]%
        {mokady2023null}
\bibfield{author}{\bibinfo{person}{Ron Mokady}, \bibinfo{person}{Amir Hertz}, \bibinfo{person}{Kfir Aberman}, \bibinfo{person}{Yael Pritch}, {and} \bibinfo{person}{Daniel Cohen-Or}.} \bibinfo{year}{2023}\natexlab{}.
\newblock \showarticletitle{Null-text Inversion for Editing Real Images using Guided Diffusion Models}. In \bibinfo{booktitle}{\emph{IEEE/CVF Conference on Computer Vision and Pattern Recognition (CVPR)}}. \bibinfo{pages}{6038--6047}.
\newblock


\bibitem[Mullender(1993)]%
        {Mullender:1993:DS(:302430}
\bibfield{editor}{\bibinfo{person}{Sape Mullender}} (Ed.). \bibinfo{year}{1993}\natexlab{}.
\newblock \bibinfo{booktitle}{\emph{Distributed systems (2nd Ed.)}}.
\newblock \bibinfo{publisher}{ACM Press/Addison-Wesley Publishing Co.}, \bibinfo{address}{New York, NY, USA}.
\newblock
\showISBNx{0-201-62427-3}


\bibitem[{National Gallery of Art}(2023)]%
        {NGA}
\bibfield{author}{\bibinfo{person}{{National Gallery of Art}}.} \bibinfo{year}{2023}\natexlab{}.
\newblock
\newblock
\urldef\tempurl%
\url{https://www.nga.gov/}
\showURL{%
\tempurl}
\newblock
\shownote{Last accessed on 2023-09-12}.


\bibitem[Nichol et~al\mbox{.}(2022)]%
        {nichol2022glide}
\bibfield{author}{\bibinfo{person}{Alex Nichol}, \bibinfo{person}{Prafulla Dhariwal}, \bibinfo{person}{Aditya Ramesh}, \bibinfo{person}{Pranav Shyam}, \bibinfo{person}{Pamela Mishkin}, \bibinfo{person}{Bob McGrew}, \bibinfo{person}{Ilya Sutskever}, {and} \bibinfo{person}{Mark Chen}.} \bibinfo{year}{2022}\natexlab{}.
\newblock \showarticletitle{{GLIDE}: Towards photorealistic image generation and editing with text-guided diffusion models}. In \bibinfo{booktitle}{\emph{International Conference on Machine Learning (ICML)}}.
\newblock


\bibitem[Nichol and Dhariwal(2021)]%
        {IDDPM}
\bibfield{author}{\bibinfo{person}{Alexander~Quinn Nichol} {and} \bibinfo{person}{Prafulla Dhariwal}.} \bibinfo{year}{2021}\natexlab{}.
\newblock \showarticletitle{Improved denoising diffusion probabilistic models}. In \bibinfo{booktitle}{\emph{International Conference on Machine Learning (ICML)}}. \bibinfo{pages}{8162--8171}.
\newblock


\bibitem[Novak(2003)]%
        {Novak03}
\bibfield{author}{\bibinfo{person}{Dave Novak}.} \bibinfo{year}{2003}\natexlab{}.
\newblock \showarticletitle{Solder man}. \bibinfo{howpublished}{Video}. In \bibinfo{booktitle}{\emph{ACM SIGGRAPH 2003 Video Review on Animation theater Program: Part I - Vol. 145 (July 27--27, 2003)}}. \bibinfo{publisher}{ACM Press}, \bibinfo{address}{New York, NY}, \bibinfo{pages}{4}.
\newblock
\urldef\tempurl%
\url{https://doi.org/99.9999/woot07-S422}
\showDOI{\tempurl}


\bibitem[Obama(2008)]%
        {Obama08}
\bibfield{author}{\bibinfo{person}{Barack Obama}.} \bibinfo{year}{2008}\natexlab{}.
\newblock \bibinfo{title}{A more perfect union}.
\newblock \bibinfo{howpublished}{Video}.
\newblock
\urldef\tempurl%
\url{http://video.google.com/videoplay?docid=6528042696351994555}
\showURL{%
Retrieved March 21, 2008 from \tempurl}


\bibitem[Park et~al\mbox{.}(2020)]%
        {park2020swapping}
\bibfield{author}{\bibinfo{person}{Taesung Park}, \bibinfo{person}{Jun-Yan Zhu}, \bibinfo{person}{Oliver Wang}, \bibinfo{person}{Jingwan Lu}, \bibinfo{person}{Eli Shechtman}, \bibinfo{person}{Alexei Efros}, {and} \bibinfo{person}{Richard Zhang}.} \bibinfo{year}{2020}\natexlab{}.
\newblock \showarticletitle{Swapping autoencoder for deep image manipulation}.
\newblock \bibinfo{journal}{\emph{Advances in Neural Information Processing Systems}}  \bibinfo{volume}{33} (\bibinfo{year}{2020}), \bibinfo{pages}{7198--7211}.
\newblock


\bibitem[Patashnik et~al\mbox{.}(2021)]%
        {StyleCLIP}
\bibfield{author}{\bibinfo{person}{Or Patashnik}, \bibinfo{person}{Zongze Wu}, \bibinfo{person}{Eli Shechtman}, \bibinfo{person}{Daniel Cohen-Or}, {and} \bibinfo{person}{Dani Lischinski}.} \bibinfo{year}{2021}\natexlab{}.
\newblock \showarticletitle{{StyleCLIP}: Text-Driven Manipulation of StyleGAN Imagery}. In \bibinfo{booktitle}{\emph{IEEE/CVF International Conference on Computer Vision (ICCV)}}. \bibinfo{pages}{2085--2094}.
\newblock


\bibitem[Petrie(1986a)]%
        {Petrie:1986:NAD:899644}
\bibfield{author}{\bibinfo{person}{Charles~J. Petrie}.} \bibinfo{year}{1986}\natexlab{a}.
\newblock \bibinfo{booktitle}{\emph{New Algorithms for Dependency-Directed Backtracking (Master's thesis)}}.
\newblock \bibinfo{type}{{T}echnical {R}eport}. \bibinfo{address}{Austin, TX, USA}.
\newblock


\bibitem[Petrie(1986b)]%
        {Petrie:1986:NAD:12345}
\bibfield{author}{\bibinfo{person}{Charles~J. Petrie}.} \bibinfo{year}{1986}\natexlab{b}.
\newblock \emph{\bibinfo{title}{New Algorithms for Dependency-Directed Backtracking (Master's thesis)}}.
\newblock \bibinfo{thesistype}{Master's\ thesis}. \bibinfo{school}{University of Texas at Austin}, \bibinfo{address}{Austin, TX, USA}.
\newblock


\bibitem[{Pexels}(2023)]%
        {pexels}
\bibfield{author}{\bibinfo{person}{{Pexels}}.} \bibinfo{year}{2023}\natexlab{}.
\newblock
\newblock
\urldef\tempurl%
\url{https://www.pexels.com}
\showURL{%
\tempurl}
\newblock
\shownote{Last accessed on 2023-09-12}.


\bibitem[Poker-Edge.Com(2006)]%
        {Poker06}
\bibfield{author}{\bibinfo{person}{Poker-Edge.Com}.} \bibinfo{year}{2006}\natexlab{}.
\newblock \bibinfo{title}{Stats and Analysis}.
\newblock
\newblock
\urldef\tempurl%
\url{http://www.poker-edge.com/stats.php}
\showURL{%
Retrieved June 7, 2006 from \tempurl}


\bibitem[Radford et~al\mbox{.}(2021)]%
        {clip}
\bibfield{author}{\bibinfo{person}{Alec Radford}, \bibinfo{person}{Jong~Wook Kim}, \bibinfo{person}{Chris Hallacy}, \bibinfo{person}{Aditya Ramesh}, \bibinfo{person}{Gabriel Goh}, \bibinfo{person}{Sandhini Agarwal}, \bibinfo{person}{Girish Sastry}, \bibinfo{person}{Amanda Askell}, \bibinfo{person}{Pamela Mishkin}, \bibinfo{person}{Jack Clark}, {et~al\mbox{.}}} \bibinfo{year}{2021}\natexlab{}.
\newblock \showarticletitle{Learning transferable visual models from natural language supervision}. In \bibinfo{booktitle}{\emph{International Conference on Machine Learning (ICML)}}. \bibinfo{pages}{8748--8763}.
\newblock


\bibitem[Ramesh et~al\mbox{.}(2022)]%
        {dalle2}
\bibfield{author}{\bibinfo{person}{Aditya Ramesh}, \bibinfo{person}{Prafulla Dhariwal}, \bibinfo{person}{Alex Nichol}, \bibinfo{person}{Casey Chu}, {and} \bibinfo{person}{Mark Chen}.} \bibinfo{year}{2022}\natexlab{}.
\newblock \showarticletitle{Hierarchical Text-Conditional Image Generation with CLIP Latents}.
\newblock \bibinfo{journal}{\emph{arXiv preprint arXiv:2204.06125}} (\bibinfo{year}{2022}).
\newblock


\bibitem[Ramesh et~al\mbox{.}(2021)]%
        {ramesh2021zero}
\bibfield{author}{\bibinfo{person}{Aditya Ramesh}, \bibinfo{person}{Mikhail Pavlov}, \bibinfo{person}{Gabriel Goh}, \bibinfo{person}{Scott Gray}, \bibinfo{person}{Chelsea Voss}, \bibinfo{person}{Alec Radford}, \bibinfo{person}{Mark Chen}, {and} \bibinfo{person}{Ilya Sutskever}.} \bibinfo{year}{2021}\natexlab{}.
\newblock \showarticletitle{Zero-shot text-to-image generation}. In \bibinfo{booktitle}{\emph{International Conference on Machine Learning (ICML)}}. PMLR, \bibinfo{pages}{8821--8831}.
\newblock


\bibitem[Rombach et~al\mbox{.}(2022)]%
        {latentdiffusion}
\bibfield{author}{\bibinfo{person}{Robin Rombach}, \bibinfo{person}{Andreas Blattmann}, \bibinfo{person}{Dominik Lorenz}, \bibinfo{person}{Patrick Esser}, {and} \bibinfo{person}{Bj{\"o}rn Ommer}.} \bibinfo{year}{2022}\natexlab{}.
\newblock \showarticletitle{High-resolution image synthesis with latent diffusion models}. In \bibinfo{booktitle}{\emph{IEEE/CVF Conference on Computer Vision and Pattern Recognition (CVPR)}}. \bibinfo{pages}{10684--10695}.
\newblock


\bibitem[Rous(2008)]%
        {Rous08}
\bibfield{author}{\bibinfo{person}{Bernard Rous}.} \bibinfo{year}{2008}\natexlab{}.
\newblock \showarticletitle{The Enabling of Digital Libraries}.
\newblock \bibinfo{journal}{\emph{Digital Libraries}} \bibinfo{volume}{12}, \bibinfo{number}{3}, Article \bibinfo{articleno}{5} (\bibinfo{year}{2008}).
\newblock
\newblock
\shownote{To appear}.


\bibitem[Ruiz et~al\mbox{.}(2023)]%
        {ruiz2022dreambooth}
\bibfield{author}{\bibinfo{person}{Nataniel Ruiz}, \bibinfo{person}{Yuanzhen Li}, \bibinfo{person}{Varun Jampani}, \bibinfo{person}{Yael Pritch}, \bibinfo{person}{Michael Rubinstein}, {and} \bibinfo{person}{Kfir Aberman}.} \bibinfo{year}{2023}\natexlab{}.
\newblock \showarticletitle{{DreamBooth}: Fine tuning text-to-image diffusion models for subject-driven generation}. In \bibinfo{booktitle}{\emph{IEEE/CVF Conference on Computer Vision and Pattern Recognition (CVPR)}}. \bibinfo{pages}{22500--22510}.
\newblock


\bibitem[Saharia et~al\mbox{.}(2022)]%
        {saharia2022photorealistic}
\bibfield{author}{\bibinfo{person}{Chitwan Saharia}, \bibinfo{person}{William Chan}, \bibinfo{person}{Saurabh Saxena}, \bibinfo{person}{Lala Li}, \bibinfo{person}{Jay Whang}, \bibinfo{person}{Emily~L Denton}, \bibinfo{person}{Kamyar Ghasemipour}, \bibinfo{person}{Raphael Gontijo~Lopes}, \bibinfo{person}{Burcu Karagol~Ayan}, \bibinfo{person}{Tim Salimans}, {et~al\mbox{.}}} \bibinfo{year}{2022}\natexlab{}.
\newblock \showarticletitle{Photorealistic text-to-image diffusion models with deep language understanding}. In \bibinfo{booktitle}{\emph{Advances in Neural Information Processing Systems (NeurIPS)}}. \bibinfo{pages}{36479--36494}.
\newblock


\bibitem[Schaldenbrand et~al\mbox{.}(2022)]%
        {StyleCLIPDraw}
\bibfield{author}{\bibinfo{person}{Peter Schaldenbrand}, \bibinfo{person}{Zhixuan Liu}, {and} \bibinfo{person}{Jean Oh}.} \bibinfo{year}{2022}\natexlab{}.
\newblock \showarticletitle{{StyleCLIPDraw}: Coupling Content and Style in Text-to-Drawing Translation}. In \bibinfo{booktitle}{\emph{International Joint Conference on Artificial Intelligence (IJCAI)}}. \bibinfo{pages}{4966--4972}.
\newblock


\bibitem[Scientist(2009)]%
        {JoeScientist001}
\bibfield{author}{\bibinfo{person}{Joseph Scientist}.} \bibinfo{year}{2009}\natexlab{}.
\newblock \bibinfo{title}{The fountain of youth}.
\newblock
\newblock
\newblock
\shownote{Patent No. 12345, Filed July 1st., 2008, Issued Aug. 9th., 2009}.


\bibitem[Singh et~al\mbox{.}(2019)]%
        {singh2019finegan}
\bibfield{author}{\bibinfo{person}{Krishna~Kumar Singh}, \bibinfo{person}{Utkarsh Ojha}, {and} \bibinfo{person}{Yong~Jae Lee}.} \bibinfo{year}{2019}\natexlab{}.
\newblock \showarticletitle{{FineGAN:} Unsupervised Hierarchical Disentanglement for Fine-Grained Object Generation and Discovery}. In \bibinfo{booktitle}{\emph{IEEE/CVF Conference on Computer Vision and Pattern Recognition (CVPR)}}. \bibinfo{pages}{6490--6499}.
\newblock


\bibitem[Smith(2010)]%
        {Smith10}
\bibfield{author}{\bibinfo{person}{Stan~W. Smith}.} \bibinfo{year}{2010}\natexlab{}.
\newblock \showarticletitle{An experiment in bibliographic mark-up: Parsing metadata for XML export}. In \bibinfo{booktitle}{\emph{Proceedings of the 3rd. annual workshop on Librarians and Computers}} \emph{(\bibinfo{series}{LAC '10}, Vol.~\bibinfo{volume}{3})}, \bibfield{editor}{\bibinfo{person}{Reginald~N. Smythe} {and} \bibinfo{person}{Alexander Noble}} (Eds.). \bibinfo{publisher}{Paparazzi Press}, \bibinfo{address}{Milan Italy}, \bibinfo{pages}{422--431}.
\newblock
\urldef\tempurl%
\url{https://doi.org/99.9999/woot07-S422}
\showDOI{\tempurl}


\bibitem[Spector(1990)]%
        {Spector90}
\bibfield{author}{\bibinfo{person}{Asad~Z. Spector}.} \bibinfo{year}{1990}\natexlab{}.
\newblock \showarticletitle{Achieving application requirements}.
\newblock In \bibinfo{booktitle}{\emph{Distributed Systems} (\bibinfo{edition}{2nd.} ed.)}, \bibfield{editor}{\bibinfo{person}{Sape Mullender}} (Ed.). \bibinfo{publisher}{ACM Press}, \bibinfo{address}{New York, NY}, \bibinfo{pages}{19--33}.
\newblock
\urldef\tempurl%
\url{https://doi.org/10.1145/90417.90738}
\showDOI{\tempurl}


\bibitem[Tao et~al\mbox{.}(2022)]%
        {Tao:2022:dfgan}
\bibfield{author}{\bibinfo{person}{Ming Tao}, \bibinfo{person}{Hao Tang}, \bibinfo{person}{Fei Wu}, \bibinfo{person}{Xiaoyuan Jing}, \bibinfo{person}{Bing-Kun Bao}, {and} \bibinfo{person}{Changsheng Xu}.} \bibinfo{year}{2022}\natexlab{}.
\newblock \showarticletitle{{DF-GAN}: A Simple and Effective Baseline for Text-to-Image Synthesis}. In \bibinfo{booktitle}{\emph{IEEE/CVF Conference on Computer Vision and Pattern Recognition (CVPR)}}. \bibinfo{pages}{16494--16504}.
\newblock


\bibitem[Tewel et~al\mbox{.}(2023)]%
        {Tewel:2023:perfusion}
\bibfield{author}{\bibinfo{person}{Yoad Tewel}, \bibinfo{person}{Rinon Gal}, \bibinfo{person}{Gal Chechik}, {and} \bibinfo{person}{Yuval Atzmon}.} \bibinfo{year}{2023}\natexlab{}.
\newblock \showarticletitle{Key-Locked Rank One Editing for Text-to-Image Personalization}. In \bibinfo{booktitle}{\emph{ACM SIGGRAPH 2023 Conference Proceedings}} (Los Angeles, CA, USA) \emph{(\bibinfo{series}{SIGGRAPH '23})}. \bibinfo{publisher}{Association for Computing Machinery}, \bibinfo{address}{New York, NY, USA}, Article \bibinfo{articleno}{12}, \bibinfo{numpages}{11}~pages.
\newblock
\showISBNx{9798400701597}


\bibitem[{The Barnes Foundation}(2023)]%
        {barnes}
\bibfield{author}{\bibinfo{person}{{The Barnes Foundation}}.} \bibinfo{year}{2023}\natexlab{}.
\newblock
\newblock
\urldef\tempurl%
\url{https://www.barnesfoundation.org/}
\showURL{%
\tempurl}
\newblock
\shownote{Last accessed on 2023-09-12}.


\bibitem[Thornburg(2001)]%
        {Thornburg01}
\bibfield{author}{\bibinfo{person}{Harry Thornburg}.} \bibinfo{year}{2001}\natexlab{}.
\newblock \bibinfo{booktitle}{\emph{Introduction to Bayesian Statistics}}.
\newblock
\urldef\tempurl%
\url{http://ccrma.stanford.edu/~jos/bayes/bayes.html}
\showURL{%
Retrieved March 2, 2005 from \tempurl}


\bibitem[Valevski et~al\mbox{.}(2023)]%
        {valevski2022unitune}
\bibfield{author}{\bibinfo{person}{Dani Valevski}, \bibinfo{person}{Matan Kalman}, \bibinfo{person}{Eyal Molad}, \bibinfo{person}{Eyal Segalis}, \bibinfo{person}{Yossi Matias}, {and} \bibinfo{person}{Yaniv Leviathan}.} \bibinfo{year}{2023}\natexlab{}.
\newblock \showarticletitle{UniTune: Text-Driven Image Editing by Fine Tuning a Diffusion Model on a Single Image}.
\newblock \bibinfo{journal}{\emph{ACM Transactions on Graphics}} \bibinfo{volume}{42}, \bibinfo{number}{4}, Article \bibinfo{articleno}{128} (\bibinfo{year}{2023}), \bibinfo{numpages}{10}~pages.
\newblock
\showISSN{0730-0301}


\bibitem[Voynov et~al\mbox{.}(2023)]%
        {voynov2023pplus}
\bibfield{author}{\bibinfo{person}{Andrey Voynov}, \bibinfo{person}{Qinghao Chu}, \bibinfo{person}{Daniel Cohen-Or}, {and} \bibinfo{person}{Kfir Aberman}.} \bibinfo{year}{2023}\natexlab{}.
\newblock \showarticletitle{$P+$: Extended Textual Conditioning in Text-to-Image Generation}.
\newblock \bibinfo{journal}{\emph{arXiv preprint arXiv:2303.09522}} (\bibinfo{year}{2023}).
\newblock


\bibitem[Wang et~al\mbox{.}(2023b)]%
        {Wang:2023:THR}
\bibfield{author}{\bibinfo{person}{Cong Wang}, \bibinfo{person}{Fan Tang}, \bibinfo{person}{Yong Zhang}, \bibinfo{person}{Tieru Wu}, {and} \bibinfo{person}{Weiming Dong}.} \bibinfo{year}{2023}\natexlab{b}.
\newblock \showarticletitle{Towards harmonized regional style transfer and manipulation for facial images}.
\newblock \bibinfo{journal}{\emph{Computational Visual Media}} \bibinfo{volume}{9}, \bibinfo{number}{2} (\bibinfo{year}{2023}), \bibinfo{pages}{351--366}.
\newblock


\bibitem[Wang et~al\mbox{.}(2023a)]%
        {Wang:2023:DFE}
\bibfield{author}{\bibinfo{person}{Zongji Wang}, \bibinfo{person}{Yunfei Liu}, {and} \bibinfo{person}{Feng Lu}.} \bibinfo{year}{2023}\natexlab{a}.
\newblock \showarticletitle{Discriminative feature encoding for intrinsic image decomposition}.
\newblock \bibinfo{journal}{\emph{Computational Visual Media}} \bibinfo{volume}{9}, \bibinfo{number}{3} (\bibinfo{year}{2023}), \bibinfo{pages}{597--618}.
\newblock


\bibitem[Wen et~al\mbox{.}(2023)]%
        {wen2023hard}
\bibfield{author}{\bibinfo{person}{Yuxin Wen}, \bibinfo{person}{Neel Jain}, \bibinfo{person}{John Kirchenbauer}, \bibinfo{person}{Micah Goldblum}, \bibinfo{person}{Jonas Geiping}, {and} \bibinfo{person}{Tom Goldstein}.} \bibinfo{year}{2023}\natexlab{}.
\newblock \showarticletitle{Hard prompts made easy: Gradient-based discrete optimization for prompt tuning and discovery}.
\newblock \bibinfo{journal}{\emph{arXiv preprint arXiv:2302.03668}} (\bibinfo{year}{2023}).
\newblock


\bibitem[Werneck et~al\mbox{.}(2000a)]%
        {Werneck:2000:FMC:351827.384253}
\bibfield{author}{\bibinfo{person}{Renato Werneck}, \bibinfo{person}{Jo\~{a}o Setubal}, {and} \bibinfo{person}{Arlindo da Conceic\~{a}o}.} \bibinfo{year}{2000}\natexlab{a}.
\newblock \showarticletitle{(new) Finding minimum congestion spanning trees}.
\newblock \bibinfo{journal}{\emph{J. Exp. Algorithmics}}  \bibinfo{volume}{5}, Article \bibinfo{articleno}{11} (\bibinfo{year}{2000}).
\newblock
\showISSN{1084-6654}
\urldef\tempurl%
\url{https://doi.org/10.1145/351827.384253}
\showDOI{\tempurl}


\bibitem[Werneck et~al\mbox{.}(2000b)]%
        {384253}
\bibfield{author}{\bibinfo{person}{Renato Werneck}, \bibinfo{person}{Jo\~{a}o Setubal}, {and} \bibinfo{person}{Arlindo da Conceic\~{a}o}.} \bibinfo{year}{2000}\natexlab{b}.
\newblock \showarticletitle{(old) Finding minimum congestion spanning trees}.
\newblock \bibinfo{journal}{\emph{J. Exp. Algorithmics}}  \bibinfo{volume}{5} (\bibinfo{year}{2000}), \bibinfo{pages}{11}.
\newblock
\showISSN{1084-6654}
\urldef\tempurl%
\url{https://doi.org/10.1145/351827.384253}
\showDOI{\tempurl}


\bibitem[Wu et~al\mbox{.}(2023)]%
        {wu2022uncovering}
\bibfield{author}{\bibinfo{person}{Qiucheng Wu}, \bibinfo{person}{Yujian Liu}, \bibinfo{person}{Handong Zhao}, \bibinfo{person}{Ajinkya Kale}, \bibinfo{person}{Trung Bui}, \bibinfo{person}{Tong Yu}, \bibinfo{person}{Zhe Lin}, \bibinfo{person}{Yang Zhang}, {and} \bibinfo{person}{Shiyu Chang}.} \bibinfo{year}{2023}\natexlab{}.
\newblock \showarticletitle{Uncovering the Disentanglement Capability in Text-to-Image Diffusion Models}. In \bibinfo{booktitle}{\emph{IEEE/CVF Conference on Computer Vision and Pattern Recognition (CVPR)}}. \bibinfo{pages}{1900--1910}.
\newblock


\bibitem[Xu et~al\mbox{.}(2018)]%
        {xu2018attngan}
\bibfield{author}{\bibinfo{person}{Tao Xu}, \bibinfo{person}{Pengchuan Zhang}, \bibinfo{person}{Qiuyuan Huang}, \bibinfo{person}{Han Zhang}, \bibinfo{person}{Zhe Gan}, \bibinfo{person}{Xiaolei Huang}, {and} \bibinfo{person}{Xiaodong He}.} \bibinfo{year}{2018}\natexlab{}.
\newblock \showarticletitle{{AttnGAN}: Fine-Grained Text to Image Generation with Attentional Generative Adversarial Networks}. In \bibinfo{booktitle}{\emph{IEEE/CVF Conference on Computer Vision and Pattern Recognition (CVPR)}}. \bibinfo{pages}{1316--1324}.
\newblock


\bibitem[Yang et~al\mbox{.}(2023a)]%
        {yang2022paint}
\bibfield{author}{\bibinfo{person}{Binxin Yang}, \bibinfo{person}{Shuyang Gu}, \bibinfo{person}{Bo Zhang}, \bibinfo{person}{Ting Zhang}, \bibinfo{person}{Xuejin Chen}, \bibinfo{person}{Xiaoyan Sun}, \bibinfo{person}{Dong Chen}, {and} \bibinfo{person}{Fang Wen}.} \bibinfo{year}{2023}\natexlab{a}.
\newblock \showarticletitle{Paint by Example: Exemplar-based Image Editing with Diffusion Models}. In \bibinfo{booktitle}{\emph{IEEE/CVF Conference on Computer Vision and Pattern Recognition (CVPR)}}. \bibinfo{pages}{18381--18391}.
\newblock


\bibitem[Yang et~al\mbox{.}(2023b)]%
        {yang2023zero}
\bibfield{author}{\bibinfo{person}{Serin Yang}, \bibinfo{person}{Hyunmin Hwang}, {and} \bibinfo{person}{Jong~Chul Ye}.} \bibinfo{year}{2023}\natexlab{b}.
\newblock \showarticletitle{Zero-Shot Contrastive Loss for Text-Guided Diffusion Image Style Transfer}.
\newblock \bibinfo{journal}{\emph{arXiv preprint arXiv:2303.08622}} (\bibinfo{year}{2023}).
\newblock


\bibitem[Ye et~al\mbox{.}(2021)]%
        {ye2021improving}
\bibfield{author}{\bibinfo{person}{Hui Ye}, \bibinfo{person}{Xiulong Yang}, \bibinfo{person}{Martin Takac}, \bibinfo{person}{Rajshekhar Sunderraman}, {and} \bibinfo{person}{Shihao Ji}.} \bibinfo{year}{2021}\natexlab{}.
\newblock \showarticletitle{Improving text-to-image synthesis using contrastive learning}.
\newblock \bibinfo{journal}{\emph{arXiv preprint arXiv:2107.02423}} (\bibinfo{year}{2021}).
\newblock


\bibitem[Yu et~al\mbox{.}(2023)]%
        {yu2022scaling}
\bibfield{author}{\bibinfo{person}{Jiahui Yu}, \bibinfo{person}{Yuanzhong Xu}, \bibinfo{person}{Jing~Yu Koh}, \bibinfo{person}{Thang Luong}, \bibinfo{person}{Gunjan Baid}, \bibinfo{person}{Zirui Wang}, \bibinfo{person}{Vijay Vasudevan}, \bibinfo{person}{Alexander Ku}, \bibinfo{person}{Yinfei Yang}, \bibinfo{person}{Burcu~Karagol Ayan}, \bibinfo{person}{Ben Hutchinson}, \bibinfo{person}{Wei Han}, \bibinfo{person}{Zarana Parekh}, \bibinfo{person}{Xin Li}, \bibinfo{person}{Han Zhang}, \bibinfo{person}{Jason Baldridge}, {and} \bibinfo{person}{Yonghui Wu}.} \bibinfo{year}{2023}\natexlab{}.
\newblock \showarticletitle{Scaling Autoregressive Models for Content-Rich Text-to-Image Generation}.
\newblock \bibinfo{journal}{\emph{Transactions on Machine Learning Research}} (\bibinfo{year}{2023}).
\newblock


\bibitem[Zhang et~al\mbox{.}(2021)]%
        {zhang2021cross}
\bibfield{author}{\bibinfo{person}{Han Zhang}, \bibinfo{person}{Jing~Yu Koh}, \bibinfo{person}{Jason Baldridge}, \bibinfo{person}{Honglak Lee}, {and} \bibinfo{person}{Yinfei Yang}.} \bibinfo{year}{2021}\natexlab{}.
\newblock \showarticletitle{Cross-Modal Contrastive Learning for Text-to-Image Generation}. In \bibinfo{booktitle}{\emph{IEEE/CVF Conference on Computer Vision and Pattern Recognition (CVPR)}}. \bibinfo{pages}{833--842}.
\newblock


\bibitem[Zhang et~al\mbox{.}(2023b)]%
        {zhang2023inst}
\bibfield{author}{\bibinfo{person}{Yuxin Zhang}, \bibinfo{person}{Nisha Huang}, \bibinfo{person}{Fan Tang}, \bibinfo{person}{Haibin Huang}, \bibinfo{person}{Chongyang Ma}, \bibinfo{person}{Weiming Dong}, {and} \bibinfo{person}{Changsheng Xu}.} \bibinfo{year}{2023}\natexlab{b}.
\newblock \showarticletitle{Inversion-Based Style Transfer with Diffusion Models}. In \bibinfo{booktitle}{\emph{IEEE/CVF Conference on Computer Vision and Pattern Recognition (CVPR)}}. \bibinfo{pages}{10146--10156}.
\newblock


\bibitem[Zhang et~al\mbox{.}(2022)]%
        {zhang2022cast}
\bibfield{author}{\bibinfo{person}{Yuxin Zhang}, \bibinfo{person}{Fan Tang}, \bibinfo{person}{Weiming Dong}, \bibinfo{person}{Haibin Huang}, \bibinfo{person}{Chongyang Ma}, \bibinfo{person}{Tong-Yee Lee}, {and} \bibinfo{person}{Changsheng Xu}.} \bibinfo{year}{2022}\natexlab{}.
\newblock \showarticletitle{Domain Enhanced Arbitrary Image Style Transfer via Contrastive Learning}. In \bibinfo{booktitle}{\emph{ACM SIGGRAPH 2022 Conference Proceedings}}. Article \bibinfo{articleno}{12}, \bibinfo{numpages}{8}~pages.
\newblock


\bibitem[Zhang et~al\mbox{.}(2023c)]%
        {Zhang:2023:UCAST}
\bibfield{author}{\bibinfo{person}{Yuxin Zhang}, \bibinfo{person}{Fan Tang}, \bibinfo{person}{Weiming Dong}, \bibinfo{person}{Haibin Huang}, \bibinfo{person}{Chongyang Ma}, \bibinfo{person}{Tong-Yee Lee}, {and} \bibinfo{person}{Changsheng Xu}.} \bibinfo{year}{2023}\natexlab{c}.
\newblock \showarticletitle{A Unified Arbitrary Style Transfer Framework via Adaptive Contrastive Learning}.
\newblock \bibinfo{journal}{\emph{ACM Transactions on Graphics}} \bibinfo{volume}{42}, \bibinfo{number}{5}, Article \bibinfo{articleno}{169} (\bibinfo{year}{2023}), \bibinfo{numpages}{16}~pages.
\newblock
\showISSN{0730-0301}


\bibitem[Zhang et~al\mbox{.}(2023a)]%
        {zhang2023sine}
\bibfield{author}{\bibinfo{person}{Zhixing Zhang}, \bibinfo{person}{Ligong Han}, \bibinfo{person}{Arnab Ghosh}, \bibinfo{person}{Dimitris Metaxas}, {and} \bibinfo{person}{Jian Ren}.} \bibinfo{year}{2023}\natexlab{a}.
\newblock \showarticletitle{SINE: SINgle Image Editing with Text-to-Image Diffusion Models}. In \bibinfo{booktitle}{\emph{IEEE/CVF Conference on Computer Vision and Pattern Recognition (CVPR)}}. \bibinfo{pages}{6027--6037}.
\newblock


\bibitem[Zhu et~al\mbox{.}(2019)]%
        {zhu2019dm}
\bibfield{author}{\bibinfo{person}{Minfeng Zhu}, \bibinfo{person}{Pingbo Pan}, \bibinfo{person}{Wei Chen}, {and} \bibinfo{person}{Yi Yang}.} \bibinfo{year}{2019}\natexlab{}.
\newblock \showarticletitle{{DM-GAN}: Dynamic Memory Generative Adversarial Networks for Text-to-Image Synthesis}. In \bibinfo{booktitle}{\emph{IEEE/CVF Conference on Computer Vision and Pattern Recognition (CVPR)}}. \bibinfo{pages}{5802--5810}.
\newblock


\end{thebibliography}

\end{document}